\pdfoutput=1
\documentclass[aps,prd,superscriptaddress,11pt,showpacs,notitlepage,nofootinbib]{revtex4}

	\usepackage{graphicx}
	\usepackage{amsmath,amssymb,mathrsfs}
	\usepackage[colorlinks=true, a4paper=true, pdfstartview=FitV,
linkcolor=blue, citecolor=blue, urlcolor=blue]{hyperref}

\usepackage{subfigure}
\usepackage{epsf}
\usepackage{epsfig}
\usepackage[usenames,dvipsnames]{xcolor}
\usepackage{bbm}
\usepackage{color}
\usepackage{comment}
\usepackage{amsmath}

\newcommand{\be}{\begin{equation}}
\newcommand{\ee}{\end{equation}}
\newcommand{\bea}{\begin{eqnarray}}
\newcommand{\eea}{\end{eqnarray}}

\newcommand{\bb}{\bibitem}

\newcommand{\eqn}{\begin{eqnarray}}
\newcommand{\eqnx}{\end{eqnarray}}

\usepackage{physics}
\numberwithin{equation}{section}

\begin{document}

\title{On decay of shock-like waves into compact oscillons}
\author{F. M. Hahne}
\thanks{CNPq Scholarship holder -- Brazil}
\author{P. Klimas}
\author{J. S. Streibel}
\affiliation{Departamento de F\'isica, Universidade Federal de Santa Catarina, Campus Trindade, 88040-900, Florian\'opolis-SC, Brazil}
\begin{abstract}
The signum-Gordon model in 1+1 dimensions possesses the exact shockwave solution with discontinuity of the field at the light cone and infinite gradient energy. The energy of a regular part of the wave inside the light cone is finite and it grows linearly with time. The initial data for such waves contain a field configuration which is null in the space and has time derivative proportional to the Dirac delta.
We study regularized initial data that lead to shock-like waves  with finite gradient energy. We found that such waves exist  in the finite time intervals and finally they decay and produce a cascade of oscillon-like structures. A pattern of the decay is very similar to the one observed in process of scattering of compact oscillons.
\end{abstract}

\maketitle
\section{Introduction}

Compact oscillons \cite{osc1} in the signum-Gordon model \cite{sg} are rather unusual field configurations. They  do not match well any of two groups formed by (quasi-) periodic  excitations of scalar fields -- oscillons and breathers.  Whereas the oscillons \cite{bogmukh, gleiser, gleiser1, roman, roman1, correa2} are quasi-periodic and slowly radiating excitations observed in non-integrable field theories, the breathers \cite{ablowitz, ablowitz2, olive, LAFWJZ} are exact and infinitely long living solutions that are present in many integrable field theories. Roughly speaking, the presence or lack of emission of the radiation from such periodic structures indicates if the model may or may not be integrable. The signum-Gordon oscillon is an exact, compact, perfectly periodic and finite energy solution. It does not radiate at all, however, when perturbed it  emits some radiation \cite{my}. Since the signum-Gordon model is a non-integrable field theory, then the existence of infinitely long living exact oscillons is a rather unusual fact. Such compact oscillons may have wider applications in theories with approximate scalling symmetry \cite{andrzej}.
It is possible due to the fact that the signum-Gordon model emerges from many field theories with non-differentaiable potentials in the limit of small amplitude excitations \cite{my, andrzej, baby, kl}. Such more general models can support topological solitons and defects \cite{shnir}.
An interesting question about the oscillons is whether the oscillons may arise in the collision process of kinks, skyrmions and other topological or non-topologigal objects.  Recently we have looked at  the process of scattering of the signum-Gordon oscillons  and emission of a radiation which, as reported in \cite{scattering}, is dominated by a huge number of smaller oscillons. Motivated by this study we have looked in more detail at the process of a collapse of the signum-Gordon shock waves which turns out to be an efficient process of production of oscillons. This issue is a main subject of the present paper. We suspect that similar phenomena may be observed in the models with approximate scalling symmetry in the limit of small amplitudes. Our study may have applications in such models and can contribute to better understanding of dynamics of  small amplitude oscillations of their fields close to vacua  \cite{adam2, andrzej}.

The signum-Gordon model  is a scalar field theory with self-interaction term which is proportional to $\phi/|\phi|$. Such a term corresponds with the potential $V=|\phi|$ which is a particular case of a wider group of so-called $\mathsf{V}$-shaped potentials.  The signum-Gordon model describes an universal behaviour of the field in vicinity of $\mathsf{V}$-shaped minima in a similar way as the Klein-Gordon equation describes dynamics of small amplitude oscillations of fields in the model with parabolic potentials around minima. According to Ref. \cite{andrzej},  a symmetry reduction in certain physical models results in limitations on values taken by new  fields and consequently in appearance of models with $\mathsf{V}$-shaped potentials. Therefore, the solutions of the signum-Gordon equation
\be
(\partial_t^2-\nabla^2)\phi(t,\vec x)+{\rm sgn}\phi(t,\vec x)=0
\ee
have universal character in the sense that they can appear as solutions (or approximations of true solutions) of many other models. For instance,  the signum-Gordon oscillons were found in the second BPS submodel of the Skyrme model \cite{andrzej, my}.

The signum-Gordon model is well behaving from physical point of view and, in particular, it admits a mechanical realization.
The potential $V=|\phi|$ is fundamentally non-linear only at $\phi=0$. Thus the signum-Gordon model shares some aspects of linear and non-linear field theories. The non-linear character is associated with the minimum of the potential {\it i.e.} it is particularly visible for small amplitude fields. On contrary to many other non-linear field theories it cannot be linearized in the limit $\phi\rightarrow 0$, which means that small amplitude perturbations are always non-linear. On the other hand, in the regions of space where the sign of the field is fixed the  signum-Gordon equation
reduces to a non-homogeneous linear wave equation.  In this paper we shall look at the real valued model in one spatial dimension.  It means that $\nabla^2\rightarrow\partial^2_x$.  In such a case there exists a general expression for solutions with fixed sign, ${\rm sgn}\phi_k=\pm1$,
\be
\phi_k(t,x)=F_k(x+t)+G_k(x-t)\pm \frac{1}{4}(x^2-t^2).\label{rozw}
\ee
where  $F_k(z)$ and $G_k(z)$ are some arbitrary functions.  Such solutions are called {\it partial solutions} and each of them has a domain corresponding with a support\footnote{Usually this support is compact.} labeled by $k$. A physical solution consists on a certain number (usually infinitely many) properly matched partial solutions. Thus linearity is rather a local property associated with individual partial solutions and it cannot be extended on solutions of the model.

The signum-Gordon model is certainly non integrable, however, it possesses quite large family of exact solutions like self-similar solutions \cite{akt, ss}, exact oscillons \cite{osc1, osc2, osc3} and shock waves \cite{akt, shock2}. This fact is related with the existence of quite general  expression for partial solutions \eqref{rozw}. In this paper we study solutions which are closely related with shock waves. The motivation for such a study came from the analysis of recent results for scattering of the signum-Gordon oscillons \cite{scattering}, where initially some shock wave-like configurations form and then they break and decay into a cascade of oscillons.

The paper is organized as follows. In Section 2 we give a short revision of shock waves in the signum-Gordon model and  present initial conditions which lead to such solutions. In Section 3 we discuss the problem of energy density and total energy of shock waves. Section 4 is devoted to presentation of numerical and analytical solutions obtained from initial configuration of the field which contains a delta-like initial profile of time derivative.

\section{The signum-Gordon shock waves}
\subsection{The solution}
The shock waves form a particular class of exact solutions of the signum-Gordon model that stand out against its other solutions by the presence of a discontinuity of the field at the light cone. This class of solutions has been proposed in Ref. \cite{akt}.  It is obtained by reduction of the signum-Gordon equation to an ordinary differential equation via ansatz
\be
\phi(t,x)=\theta(-z)W(z),\qquad {\rm where}\qquad z=\frac{1}{4}(x^2-t^2).
\ee
The function $W(z)$ obeys the  equation
\be
zW''(z)+W'(z)={\rm sgn}(W(z))\nonumber
\ee
 and it consists of infinitely many partial solutions $W_k(z)$, $k\in{\mathbb Z}$ matched at points $-a_k$. Each partial solution satisfies equation
\be
zW_k''(z)+W_k'(z)=(-1)^k\label{ordinaryeq}
\ee
and the matching conditions $W_k(-a_k)=0=W_{k+1}(-a_k)$ and $W'_k(-a_k)=W'_{k+1}(-a_k)$. Imposing the condition $W_k(-a_k)=0$ we get the partial solutions in the form
\[
W_k(z)=(-1)^k\left(z+a_k+b_k\ln\frac{|z|}{a_k}\right).
\]
Then, imposing the remaining conditions we get some restrictions on the coefficients $a_k$ and $b_k$, namely
\be
\frac{b_{k+1}}{a_k}=2-\frac{b_k}{a_k},\qquad {\rm and}\qquad\frac{a_{k+1}}{a_k}=1+\frac{b_{k+1}}{a_k}\ln \frac{a_{k+1}}{a_k}.\label{rekurencja}
\ee
\begin{figure}[h!]
\centering
\subfigure[$\quad W_k(x)$]{\includegraphics[width=0.43\textwidth,height=0.28\textwidth, angle =0]{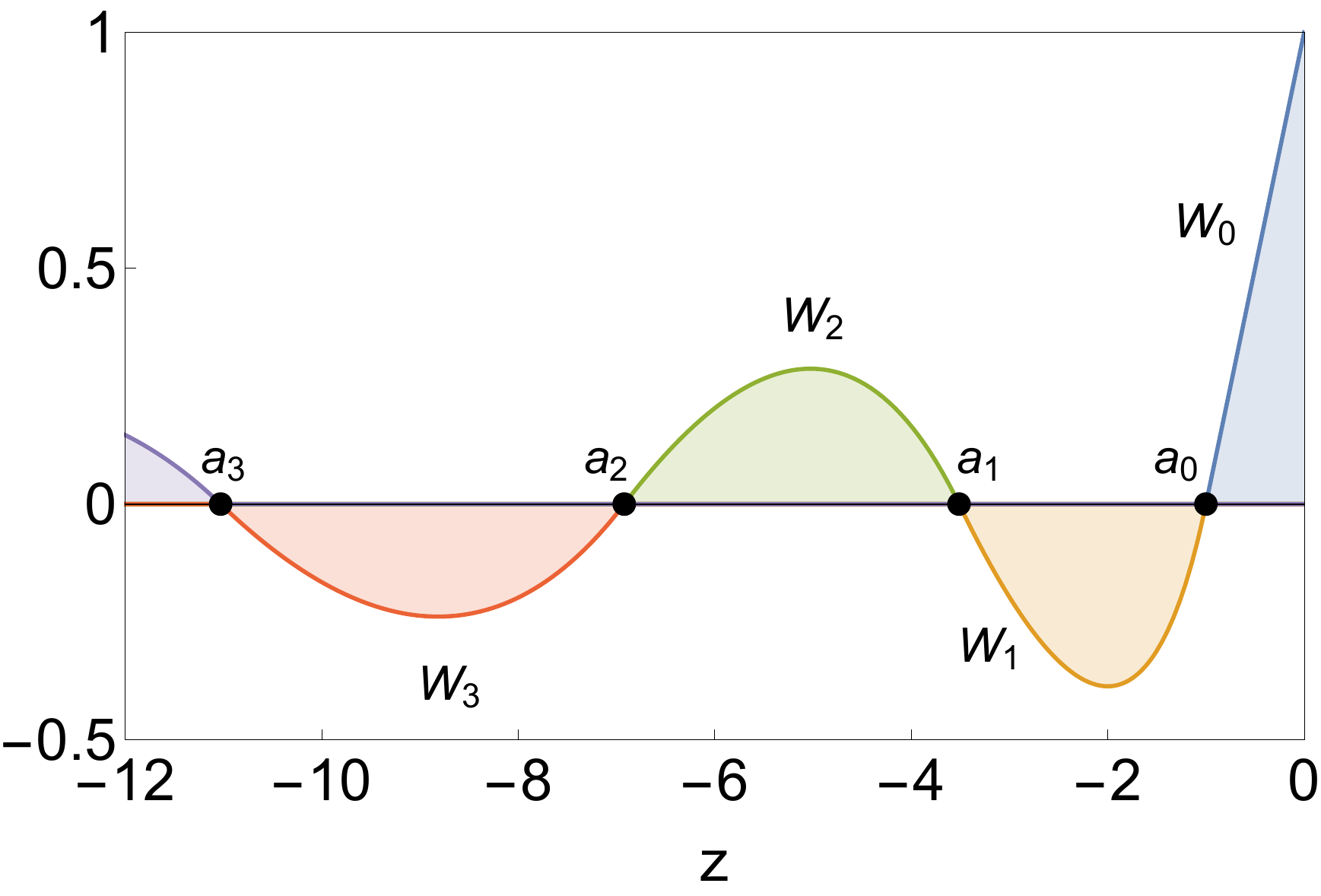}}
\hskip0.5cm
\subfigure[$\quad \phi_k(t=6.5,x)$]{\includegraphics[width=0.43\textwidth,height=0.28\textwidth, angle =0]{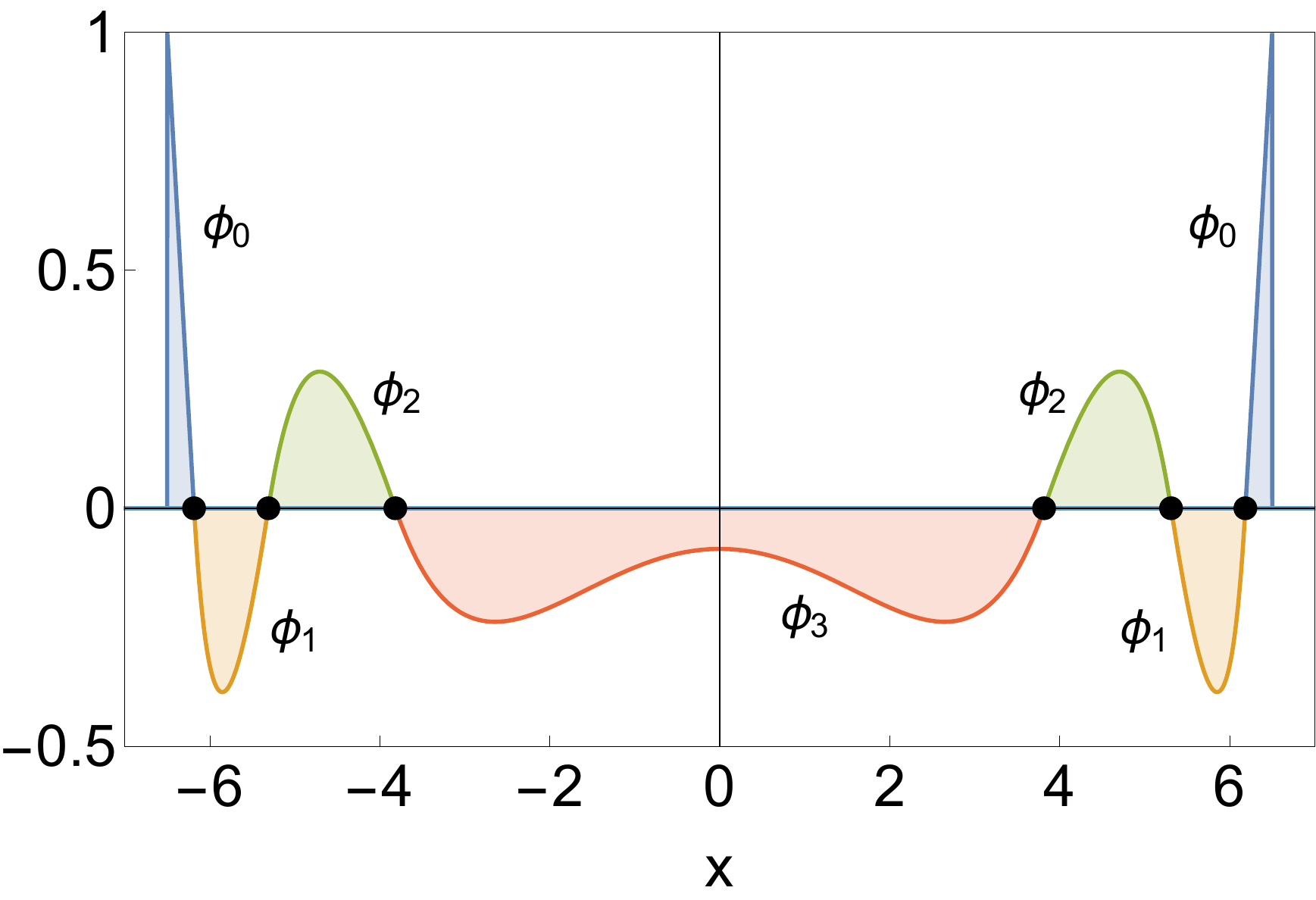}}
\caption{The exact shock wave solution for $a_0=1$. }
\label{fig:shock}
\end{figure}
Note that $b_0$ must vanish in order to avoid singularity of the logarithm at $z=0$. The first zero $a_0$ can be chosen as a free parameter which determines values of all other constants via recurrence relations \eqref{rekurencja}. In particular, it holds $b_1=2a_0$. In terms of new variables $\alpha_{k+1}:=\frac{1}{2}\frac{b_{k+1}}{a_k}$ and $y_{k+1}:=\frac{a_{k+1}}{a_k}$ the relations \eqref{rekurencja} take the form
\be
\alpha_{k+1}=1-\frac{\alpha_k}{y_k},\qquad{\rm and}\qquad y_{k+1}=1+2\alpha_{k+1}\ln y_{k+1}\label{rekurencja2}
\ee
where $\alpha_1=1$. It follows from \eqref{rekurencja2} that $\alpha_{k+1}$ is determined by $a_k$ and $b_k$.  Then, solving numerically the second equation of \eqref{rekurencja2} one gets $y_{k+1}$ and thus both coefficients $a_{k+1}$ and $b_{k+1}$ can be determined. We shall skip discussion of solution of the recurrence relations because it is given in previous papers. In Fig.\ref{fig:shock}(a) we present the first few partial solutions $W_k(z)$ and in Fig.\ref{fig:shock}(b) we present the shock wave at $t=6.5$, which consists of four partial solutions.
\begin{figure}[h!]
\centering
{\includegraphics[width=0.6\textwidth,height=0.3\textwidth, angle =0]{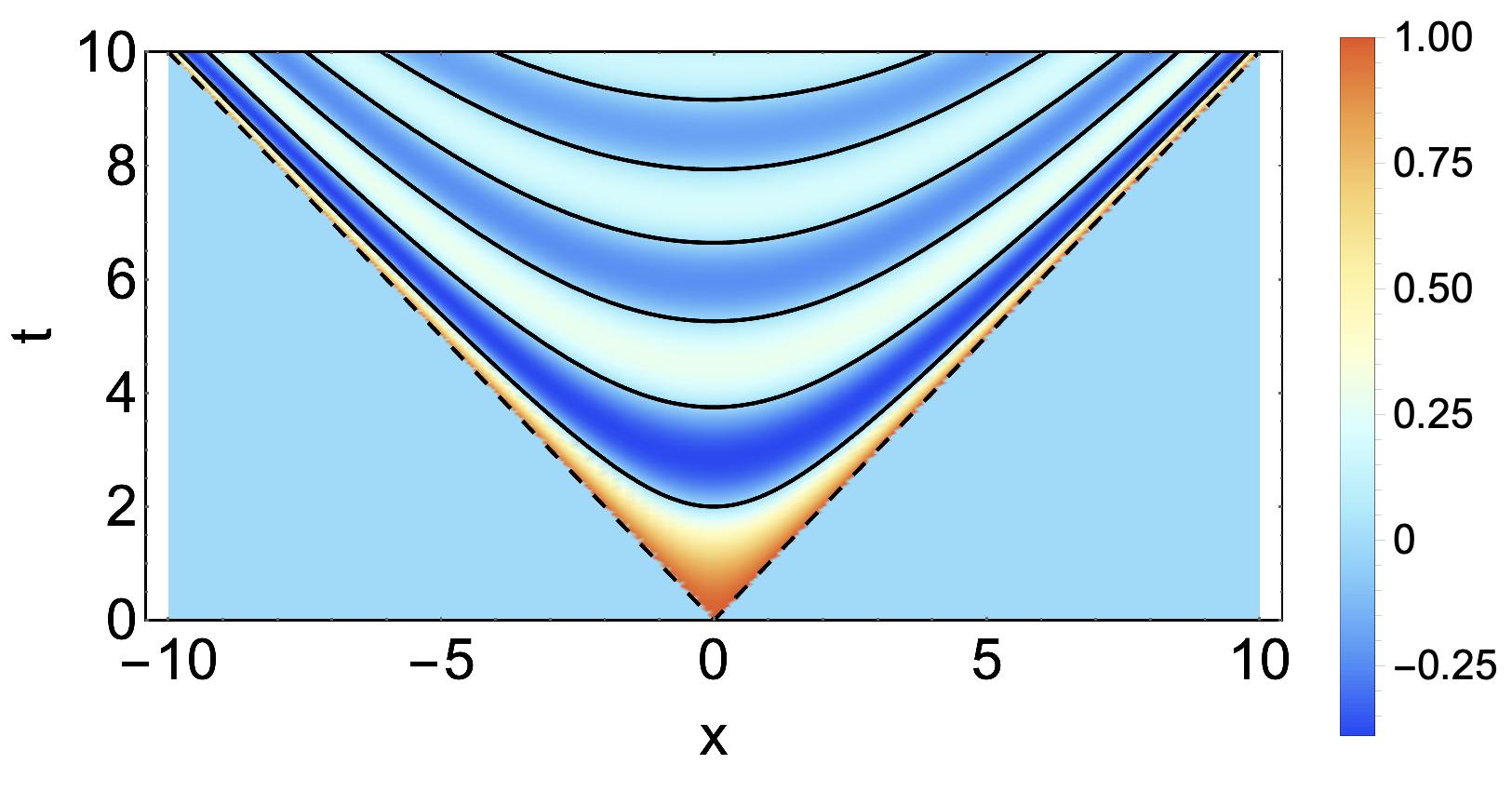}}
\caption{The exact shock wave solution for $a_0=1$ at the spacetime diagram.}
\label{fig:shock1}
\end{figure}

In Fig.\ref{fig:shock1} we show the shock wave on the Minkowski diagram. The zeros of the field $\phi(t,x)$ correspond with hyperbolas
\begin{equation}
	x_k(t)=\pm\sqrt{t^2-4a_k}.
\end{equation}

In contrast to the case of compact oscillons or self-similar solutions, there is very little known about the role of shock waves in the dynamics of more complex field configurations of in the signum-Gordon model. We have not seen such solutions in numerical simulations of interaction between kinks and perturbations of exact oscillons. However, this situation has changed after getting the first results of  scattering of high speed oscillons, see Ref.\cite{scattering}. We have found that such oscillons produce radiation that mainly consists of perturbed oscillons.  The oscillons do not appear immediately after the collision. The numerical study shows that they are generated in the process of breaking of the wave field configuration. We have confirmed that these waves are related to shock waves by comparing their zeros with zeros of exact shock waves. It raised the question about  initial conditions that support such shock wave-like configurations.   We shall answer this question in the subsequent section.

\subsection{The initial problem}

An important question we would like to answer now is a relation between discontinuity of the field at the light cone and the initial conditions which would produce such a discontinuity. This problem has not been answered in the original papers addressed to construction of shock waves in the signum-Gordon model and its modifications. The analysis of the shock wave solution allows to conclude that in the limit $t\rightarrow 0+$ the scalar field $\phi$ and its first derivative $\partial_t\phi$ vanish everywhere except the point $x=0$. It suggests the possible choice of  initial configuration of the field $\phi(0,x)=0$ everywhere and $\partial_t\phi(t,x)|_{t=0}$ proportional to Dirac delta $\delta(x)$.  The problem of discontinuity of the field at the light cone can be studied without necessity of solving the complete signum-Gordon equation.

\subsubsection{The initial problem for wave equation}
Looking at the problem of appearance of discontinuities of the field, it is enough to look at a simpler problem involving the wave equation in (1+1) dimensions $(\partial_t^2-\partial_x^2)\phi(t,x)=0$. There is another advantage associated with this simplification -- we have a general solution of the initial problem given by the d'Alembert formula. Let $\phi(0,x)=f(x)$ and $\partial_t\phi(t,x)|_{t=0}=g(x)$ be an initial field configuration, then the wave equation in one spatial dimension has solution
\be
\phi(t,x)=\frac{1}{2}\Big(f(x+t)+f(x-t)\Big)+\frac{1}{2}\int_{x-t}^{x+t}d\xi g(\xi)\label{soldal}
\ee
Taking a particular initial conditions which are suitable for our considerations
\be
\phi(0,x)=f(x)=0,\qquad\qquad \partial_t\phi(t,x)|_{t=0}=g(x)=a\delta(x)\label{incond}
\ee
where $a=const$ we find
\be
\phi(t,x)=\frac{a}{2}\Big(\theta(x+t)-\theta(x-t)\Big)=\left\{\begin{array}{ccc}
\frac{a}{2}&{\rm for}&|x|<t\\
0&{\rm for}&|x|>t
\end{array}\right..\label{soltheta}
\ee
Thus the solution $\phi(t,x)$ is equal to $a/2$ inside the light cone and it vanishes outside the light cone. Clearly, the field possesses discontinuity at the light cone $x=\pm t$. The same behavior at the light cone is expected for the signum-Gordon equation. However, in the signum-Gordon model the sign term modifies the solution inside the light cone, which leads to observed oscillations.

\subsubsection{Regular initial data}
The presence of discontinuity of the field has some unpleasant consequences. This is pretty clear when looking at the problem of energy associated with such field configurations. To see it, we look again at the problem restricted to solutions of the wave equation in one spatial dimension.  The energy of a solution is given by the expression
\be
E=\frac{1}{2}\int_{-\infty}^{\infty}dx\Big((\partial_t\phi)^2+(\partial_x\phi)^2\Big).
\ee
Looking at partial derivatives of solution \eqref{soltheta} one gets
\begin{align}
\partial_t\phi(t,x)&=\frac{a}{2}(\delta(x+t)+\delta(x-t)),\\ \partial_x\phi(t,x)&=\frac{a}{2}(\delta(x+t)-\delta(x-t)).
\end{align}
It rises the problem about expression ``$(\delta(x))^2$'' which is meaningless as a distribution. Thus we see that this question is related to the problem of ``multiplication of generalized functions".

In order to overcome this problem we shall replace the delta distribution $\delta(x)$ in  \eqref{incond} by a classical function. There are many classical functions $\delta_{\epsilon}(x)$ which  tend to the Dirac delta when $\epsilon\rightarrow 0$. This limit has the following sence
\[
\lim_{\epsilon\rightarrow 0} \int_{-\infty}^{\infty}dx \delta_{\epsilon}(x)\varphi(x)=\varphi(0)\equiv \int_{-\infty}^{\infty}dx \delta(x)\varphi(x),
\]
where $\varphi(x)$ is a test function.
 \begin{figure}[h!]
\centering
\subfigure[$\quad \delta_{\epsilon}(x)$ - Gaussian]{\includegraphics[width=0.3\textwidth,height=0.2\textwidth, angle =0]{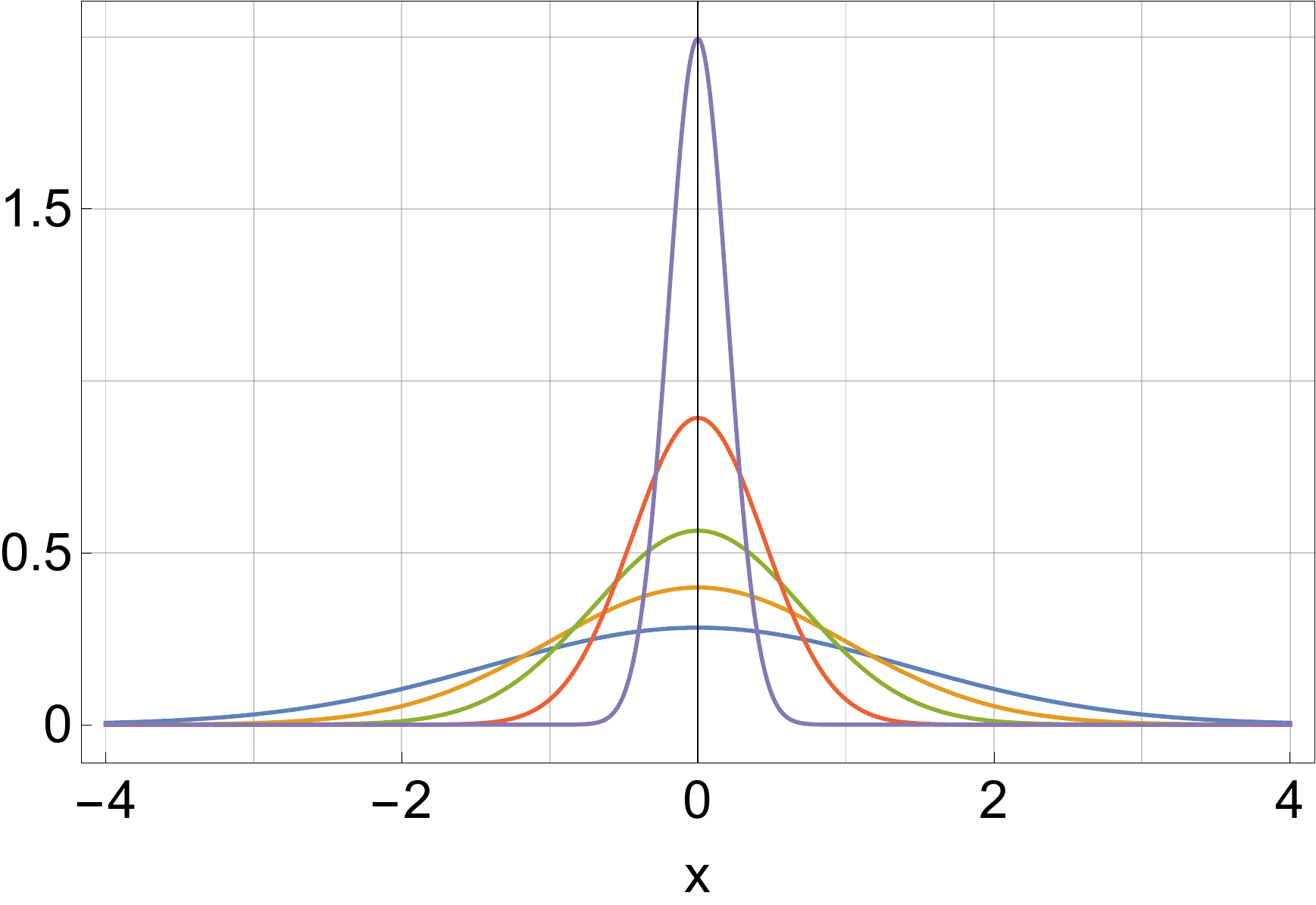}}
\hskip 0.2cm
\subfigure[$\quad \phi(t=3,x),\quad a=1$]{\includegraphics[width=0.3\textwidth,height=0.2\textwidth, angle =0]{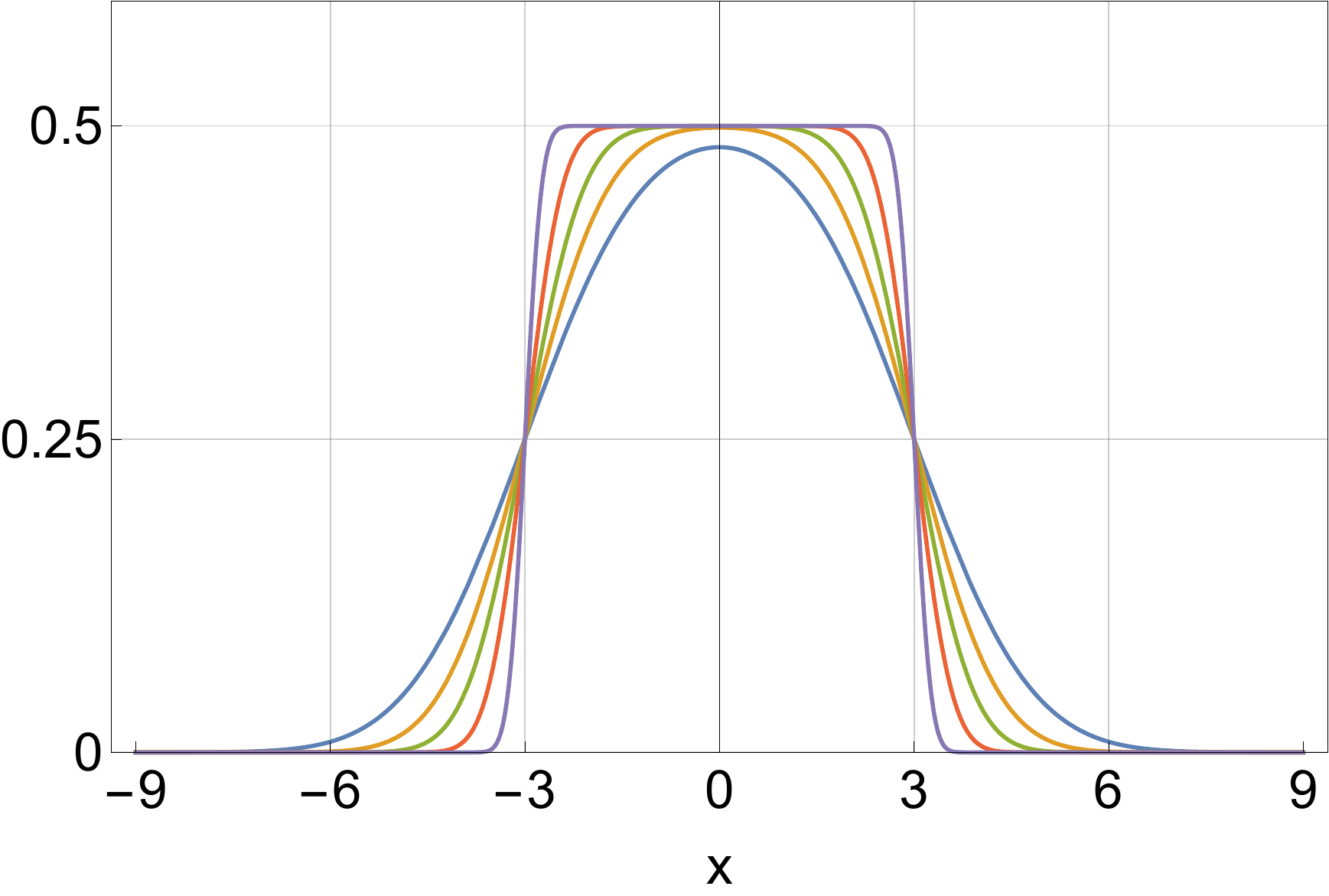}}
\hskip 0.2cm
\subfigure[$\quad u(t=3,x),\quad a=1$]{\includegraphics[width=0.3\textwidth,height=0.2\textwidth, angle =0]{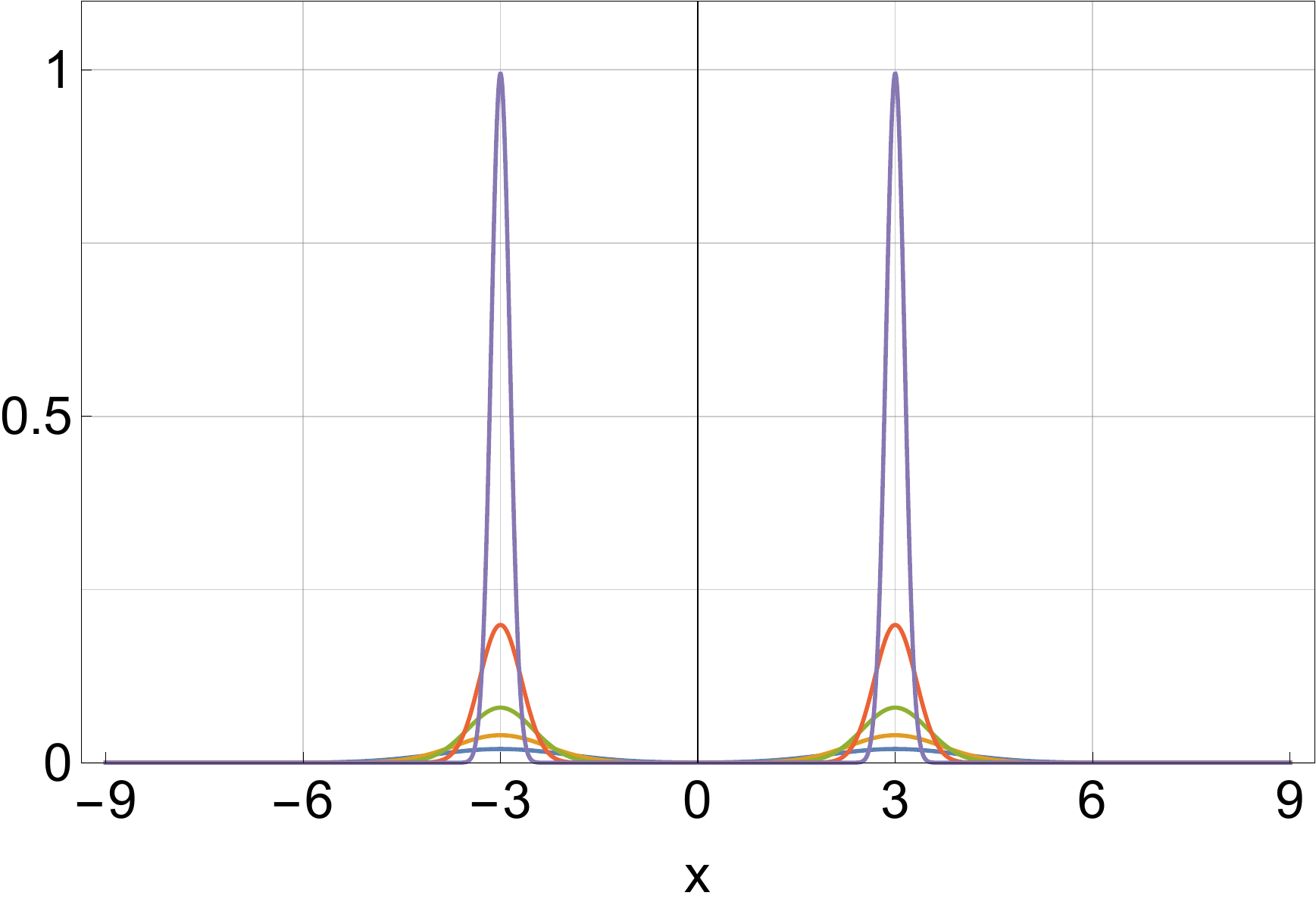}}
\caption{(a) Function $\delta_\epsilon(x)$ and (b) solution $\phi(t,x)$ and (c) energy density  for $\epsilon=\{1,\frac{1}{2},\frac{1}{4},\frac{1}{10},\frac{1}{50}\}$.}
\label{fig:dalembert}
\end{figure}
\begin{figure}[h!]
\centering
\subfigure[$\quad \delta_{\epsilon}(x)$ - triangular]{\includegraphics[width=0.3\textwidth,height=0.2\textwidth, angle =0]{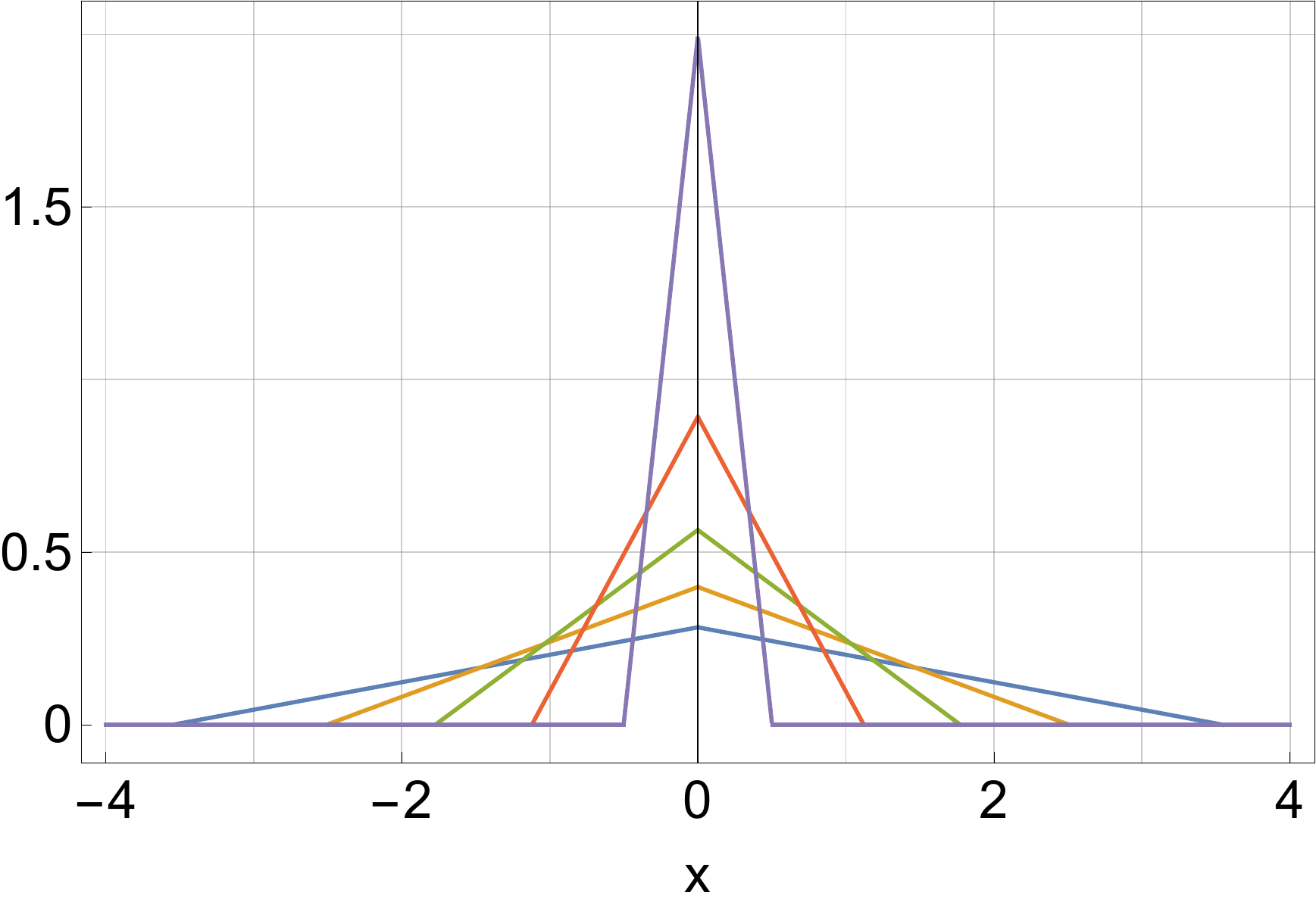}}
\hskip 0.2cm
\subfigure[$\quad \phi(t=3,x),\quad a=1$]{\includegraphics[width=0.3\textwidth,height=0.2\textwidth, angle =0]{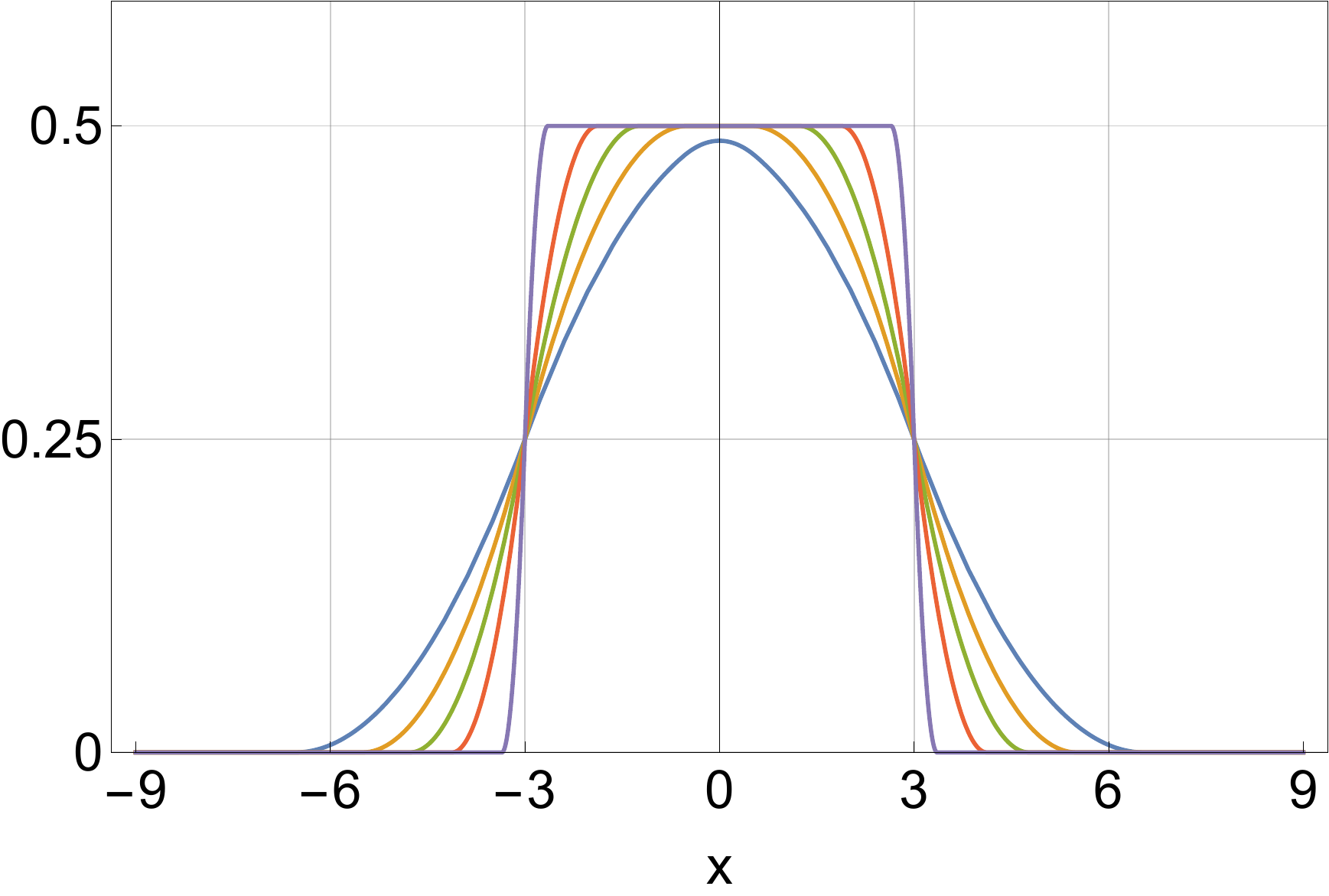}}
\hskip 0.2cm
\subfigure[$\quad u(t=3,x),\quad a=1$]{\includegraphics[width=0.3\textwidth,height=0.2\textwidth, angle =0]{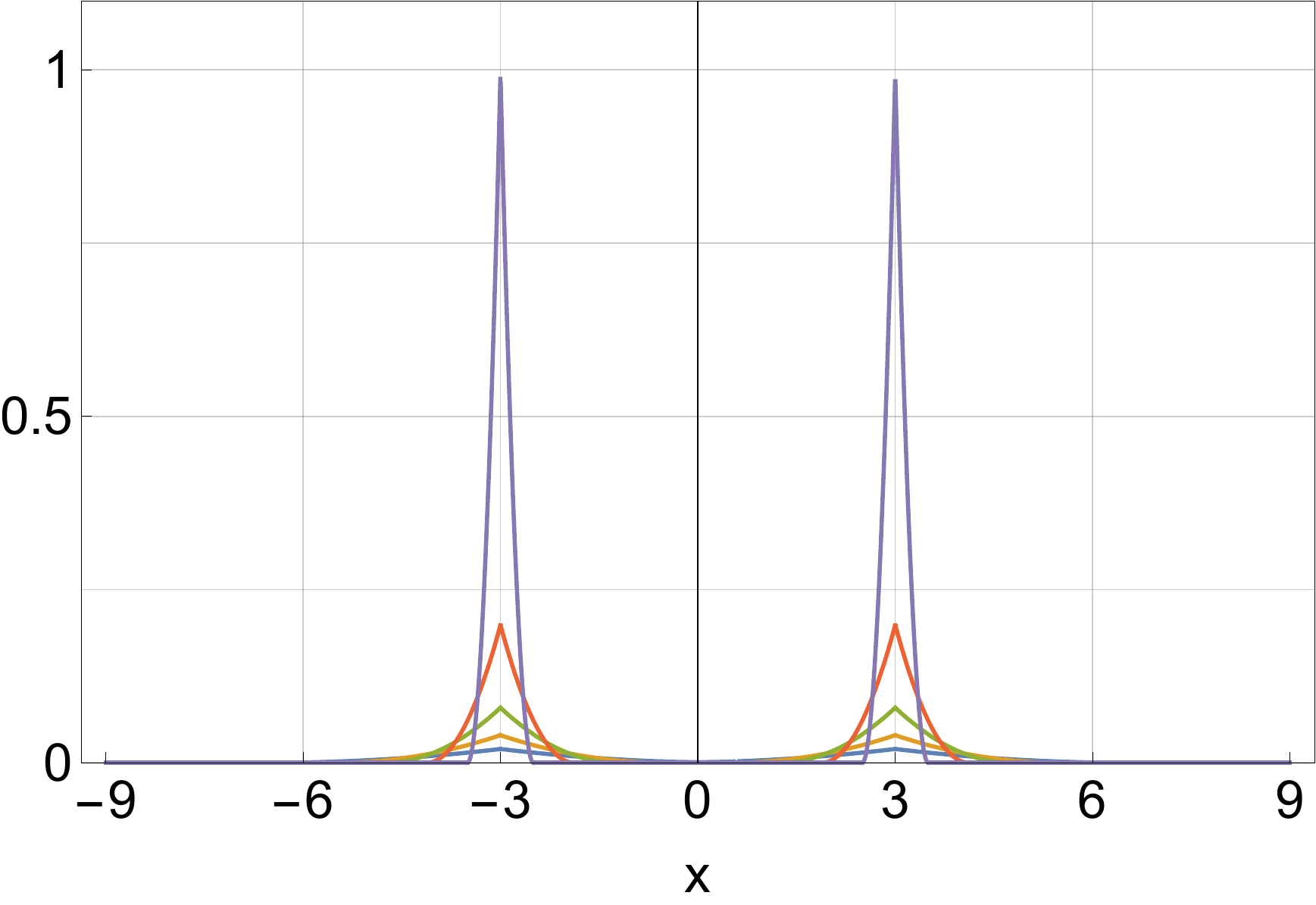}}
\caption{(a) Function $\delta_\epsilon(x)$ and (b) solution $\phi(t,x)$ and (c) energy density  for $\epsilon=\{1,\frac{1}{2},\frac{1}{4},\frac{1}{10},\frac{1}{50}\}$.}
\label{fig:dalembert1}
\end{figure}

We  shall consider two such functions in our analysis, namely a Gaussian function
\begin{equation}
\delta_{\epsilon}(x)=\frac{1}{2\sqrt{\pi\epsilon}}e^{-\frac{x^2}{4\epsilon}}\label{gauss}
\end{equation}
and a triangular function with compact support
\begin{equation}
\delta_{\epsilon}(x)=\frac{1}{4\pi\epsilon}\Big[(2\sqrt{\pi \epsilon}+x)\theta(2\sqrt{\pi \epsilon}+x)\theta(-x)+(2\sqrt{\pi \epsilon}-x)\theta(2\sqrt{\pi \epsilon}-x)\theta(x)\Big].\label{triangle}
\end{equation}

We have chosen two different class of functions $\delta_{\epsilon}$ in order to see to which extent solutions of the wave equation depend on the choice.
Functions \eqref{gauss} and \eqref{triangle} satisfy
\[
\int_{-\infty}^{\infty}dx\,\delta_{\epsilon}(x)=1,\qquad\qquad  \delta_{\epsilon}(0)=\frac{1}{2\sqrt{\pi\epsilon}}.
\]
The triangular function $\delta_{\epsilon}$  has been normalized in the way that it takes the same value at the center $x=0$ as the Gaussian function. The plots corresponding with \eqref{gauss} are shown in Fig.\ref{fig:dalembert}(a) and those corresponding with \eqref{triangle} in Fig.\ref{fig:dalembert1}(a).

\begin{figure}[h!]
\centering
{\includegraphics[width=0.5\textwidth,height=0.35\textwidth, angle =0]{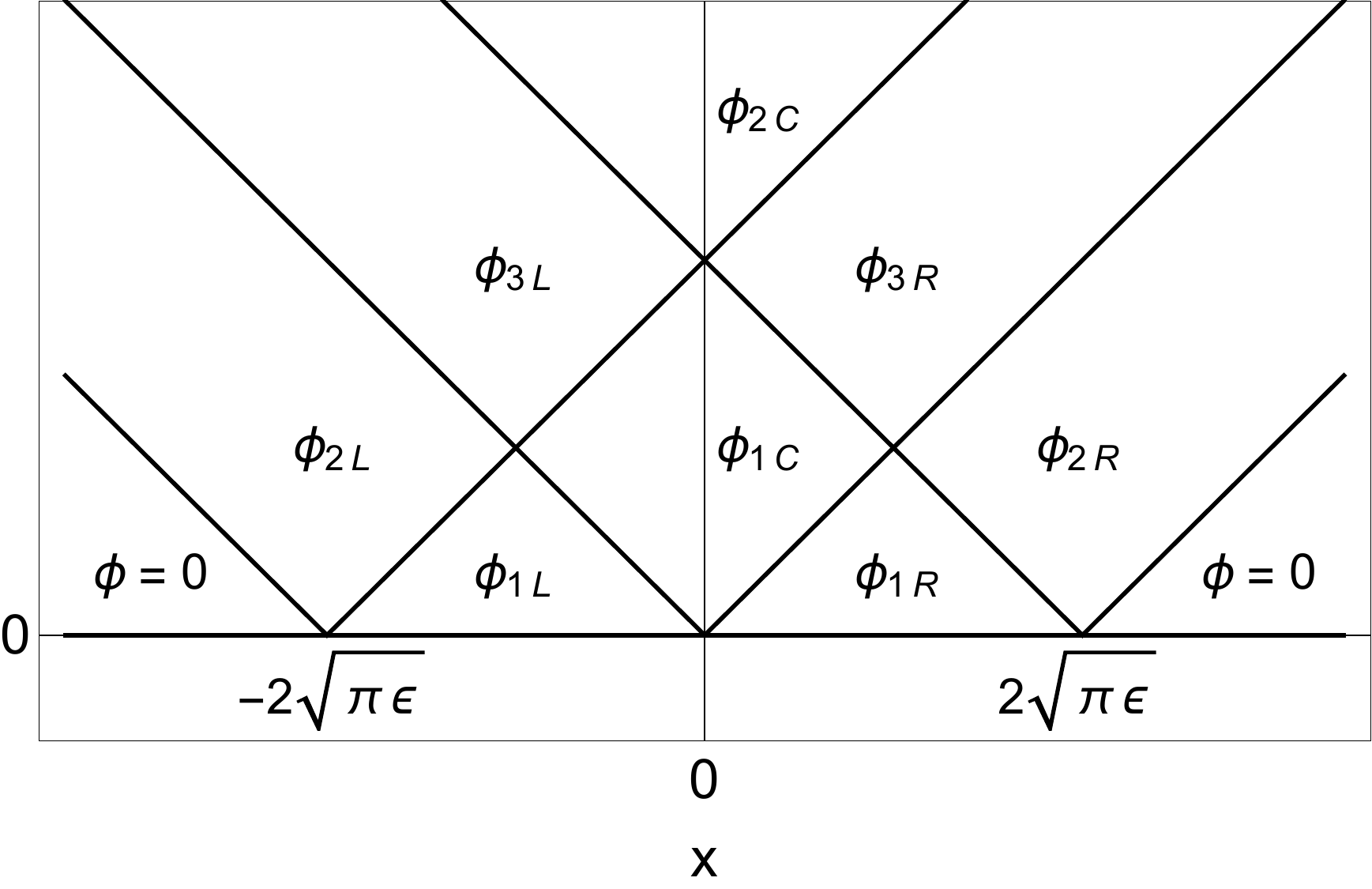}}
\caption{Partial solutions for triangle-shape initial profile of $\partial_t\phi$.}
\label{fig:partial}
\end{figure}

The d'Alembert formula \eqref{soldal} allows for obtaining an exact solution for $f(x)=0$ and $g(x)=a\delta_{\epsilon}(x)$. When $\delta_{\epsilon}$ is a Gaussian function one gets expression
\be
\phi(t,x)=\frac{a}{4}\Big(\erf\Big[\frac{x+t}{2\sqrt{\epsilon}}\Big]-\erf\Big[\frac{x-t}{2\sqrt{\epsilon}}\Big]\Big)\label{solgauss}
\ee
where $\erf(z)$ stands for the error function\footnote{$\erf(z):=\frac{2}{\sqrt{\pi}}\int_0^z du e^{-u^2}=\frac{2}{\sqrt{\pi}}\sum_{n=0}^{\infty}\frac{(-1)^n z^{2n+1}}{n!(2n+1)}.$}.  Fig.\ref{fig:dalembert}(b) shows a few solutions $\phi(t,x)$ at $t=3$ for different values of the parameter $\epsilon$. In the limit $\epsilon\rightarrow 0$ solution \eqref{solgauss}  tends to \eqref{soltheta}.
In the case of initial data given by \eqref{triangle} one can also obtain exact solution, however, this solution is technically more complex because it contains different expressions in different regions of spacetime. The partial solutions vanish outside their supports. In Fig.\ref{fig:partial} we show regions where are localized supports of partial solutions. The partial solutions that form the solution of the problem are given by expressions
\begin{align}
\phi_{1L}(t,x)&=\frac{a}{4\pi\epsilon}(2\sqrt{\pi\epsilon}+x)t,\label{phiweL1}\\
\phi_{2L}(t,x)&=\frac{a}{4\pi\epsilon}\left(\frac{x+t}{2}+\sqrt{\pi\epsilon}\right)^2,\label{phiweL2}\\
\phi_{3L}(t,x)&=\frac{a}{4\pi\epsilon}\left(\pi\epsilon+\sqrt{\pi\epsilon}(x+t)-\frac{(x+t)^2}{4}\right),\label{phiweL3}\\
\phi_{1C}(t,x)&=\frac{a}{4\pi\epsilon}\left(2\sqrt{\pi\epsilon}t-\frac{x^2+t^2}{2}\right),\label{phiweC1}\\
\phi_{2C}(t,x)&=\frac{a}{2},\label{phiweC2}\\
\phi_{kR}(t,x)&=\phi_{kL}(t,-x)\qquad \text{for}\qquad k=1,2,3.\nonumber
\end{align}

Note that all expressions presented above are given by polynomials of degree no higher than two. Fig.\ref{fig:dalembert1}(b)
shows the profile of the solution $\phi(t,x)$ at $t=3$ for various values of $\epsilon$. In the case of initial data given by triangle-shape of initial profile of $\partial_t\phi$ the solution is compact.

Finally, one can look at the energy density and total energy of such solutions. Since the solutions $\phi(t,x)$ is given by classical functions then there is no problem with taking square of its derivatives. In the case of solution \eqref{solgauss} obtained for Gaussian initial data one gets the following expression for a total energy of the field
\be
E=\int_{-\infty}^{\infty}dx\,u(t,x)=\frac{a^2}{16\pi\epsilon}\int_{-\infty}^{\infty}dx(1+e^{\frac{2tx}{\epsilon}})e^{-\frac{(x+t)^2}{2\epsilon}}=\frac{a^2}{4\sqrt{2\pi\epsilon}}.\label{E1}
\ee
The energy density shown in Fig.\ref{fig:dalembert}(c) is finite everywhere. Also the total energy of the field is finite, however, it depends on $\epsilon$. In the limit $\epsilon\rightarrow 0$ the energy tends to infinity as $\epsilon^{-1/2}$ what is associated with formation of discontinuities of the field and increasing of the gradient energy.

In the case of triangle-shape profile \eqref{triangle} we find that the energy density is a  finite function that consists on patches
\begin{align}
u_{1L}(t,x)&=\frac{a^2}{32\pi^2\epsilon^2}\Big[t^2 + (x + 2 \sqrt{\pi\epsilon})^2\Big],\nonumber\\
u_{2L}(t,x)&=\frac{a^2}{64\pi^2\epsilon^2}\Big[t + x + 2 \sqrt{\pi\epsilon}\Big]^2,\nonumber\\
u_{3L}(t,x)&=\frac{a^2}{64\pi^2\epsilon^2}\Big[t + x - 2 \sqrt{\pi\epsilon}\Big]^2,\nonumber\\
u_{1C}(t,x)&=\frac{a^2}{32\pi^2\epsilon^2}\Big[x^2 + (t - 2 \sqrt{\pi\epsilon})^2\Big],\nonumber\\
u_{2C}(t,x)&=0,\nonumber\\
u_{kR}(t,x)&=u_{kL}(t,-x)\qquad\text{where}\qquad k=1,2,3.\nonumber
\end{align}
The plot of this energy density is shown in Fig.\ref{fig:dalembert1}(c). In similarity to the Gaussian case the energy density is a finite function for $\epsilon\neq 0$. The only difference is that now the energy density vanishes explicitly outside two compact supports localized around the light cone of the event $(0,0)$. A total energy in this case reads
\be
E=\int_{-2\sqrt{\pi\epsilon}}^{2\sqrt{\pi\epsilon}}dx\, u(t,x) =\frac{a^2}{6\sqrt{\pi\epsilon}}.\label{E2}
\ee
It has exactly the same functional behavior as for the Gaussian case.

\section{The energy of a shock wave}

A fundamental difference between the wave equation and the signum-Gordon equation is the presence of the self interaction described by potential $V(\phi)=|\phi|$  which gives rise to the term ${\rm sgn}(\phi)$ in the field equation. It means that each non-vanishing field configuration has also some potential energy. Consequently,  the wave on an open segment $-t<x<t$ has some finite energy. This energy consists on three elements: kinetic, gradient and potential energy. Of course, in similarity to the solution of the d'Alembert equation, there is also an infinite gradient energy due to discontinuities at $x=\pm t$. The gradient energy stored in discontinuity of the field at the light cone is essential for the existence of shock waves. The discontinuities of the field constitute a sort of energy reservoir feeding  the regular wave in the central region inside the future light cone of the event $(0,0)$. In order to make this statement clearer, we look in more detail to the wave on an open segment $-t<x<t$, especially to the energies of its partial solutions.

A general solution describing a shock wave is a sum\footnote{Each partial solution vanishes outside its own support.} of partial solutions $\phi_k(t,x)=W_k(z)$
\be
\phi(t,x)=\sum_{k=0}^{\infty}\phi_k(t,x)\label{sumpart}
\ee
where $W_{k}(z)\equiv 0$ outside the support $a_{k}\le z\le a_{k-1}$. At the instant of time
\be
t_{k-1}=2\sqrt{a_{k-1}},\label{tk}
\ee
the partial solution $\phi_{k-1}$ splits into two disjoint parts and there appears the solution $\phi_k(t,x)$. The solution $\phi_k(t,x)$ exists on the support
\begin{eqnarray}
{\rm supp}_{\phi_k}(t):=\left\{\begin{array}{rcr}
|x|\le c_{k-1}(t)&\quad{\rm for}\quad& t_{k-1}\le t\le t_k\\
c_k(t)\le|x|\le c_{k-1}(t)&\quad{\rm for}\quad&t\ge t_k
\end{array}\right.
\end{eqnarray}
where
\be
c_k(t):=\sqrt{t^2-4a_k}\label{ck}
\ee
have interpretation of positive zeros of the wave.
The values of $a_k$ and $t_k$ for $a_0=1$ are given in Table \ref{tab1}.  The function \eqref{sumpart} is a ${\cal C}^1$ class at $x\in(-t,t)$.
\begin{table}[h!]
\begin{tabular}{llllllll}
\hline
\multicolumn{1}{|l|}{}  & \multicolumn{1}{c|}{$k=0$} & \multicolumn{1}{c|}{$k=1$} & \multicolumn{1}{c|}{$k=2$} & \multicolumn{1}{c|}{$k=3$} & \multicolumn{1}{c|}{$k=4$} & \multicolumn{1}{c|}{$k=5$} & \multicolumn{1}{c|}{$k=6$} \\ \hline
\multicolumn{1}{|c|}{$a_k$} & \multicolumn{1}{c|}{$1.00$}  & \multicolumn{1}{c|}{$3.51$}  & \multicolumn{1}{c|}{$6.92$} & \multicolumn{1}{c|}{$11.03$} & \multicolumn{1}{c|}{$15.73$} & \multicolumn{1}{c|}{$20.96$} & \multicolumn{1}{c|}{$26.67$} \\ \hline
\multicolumn{1}{|c|}{$t_k$} & \multicolumn{1}{c|}{$2.00$}  & \multicolumn{1}{c|}{$3.75$}  & \multicolumn{1}{c|}{$5.26$} & \multicolumn{1}{c|}{$6.64$} & \multicolumn{1}{c|}{$7.93$} & \multicolumn{1}{c|}{$9.16$} & \multicolumn{1}{c|}{$10.33$} \\ \hline
                        &                        &                        &                       &                       &                       &                       &
\end{tabular}
\caption{Zeros $a_k$ of $W(z)$ and instants of time $t_k=2\sqrt{a_k}$ at which appear solutions $\phi_{k+1}(t,x)$.}\label{tab1}
\end{table}

\begin{figure}[h!]
\centering
\subfigure[$\quad \frac{1}{2}(\partial_t\phi_k)^2+\frac{1}{2}(\partial_x\phi_k)^2,\quad k=0,1,\ldots,6$]{\includegraphics[width=0.45\textwidth,height=0.25\textwidth, angle =0]{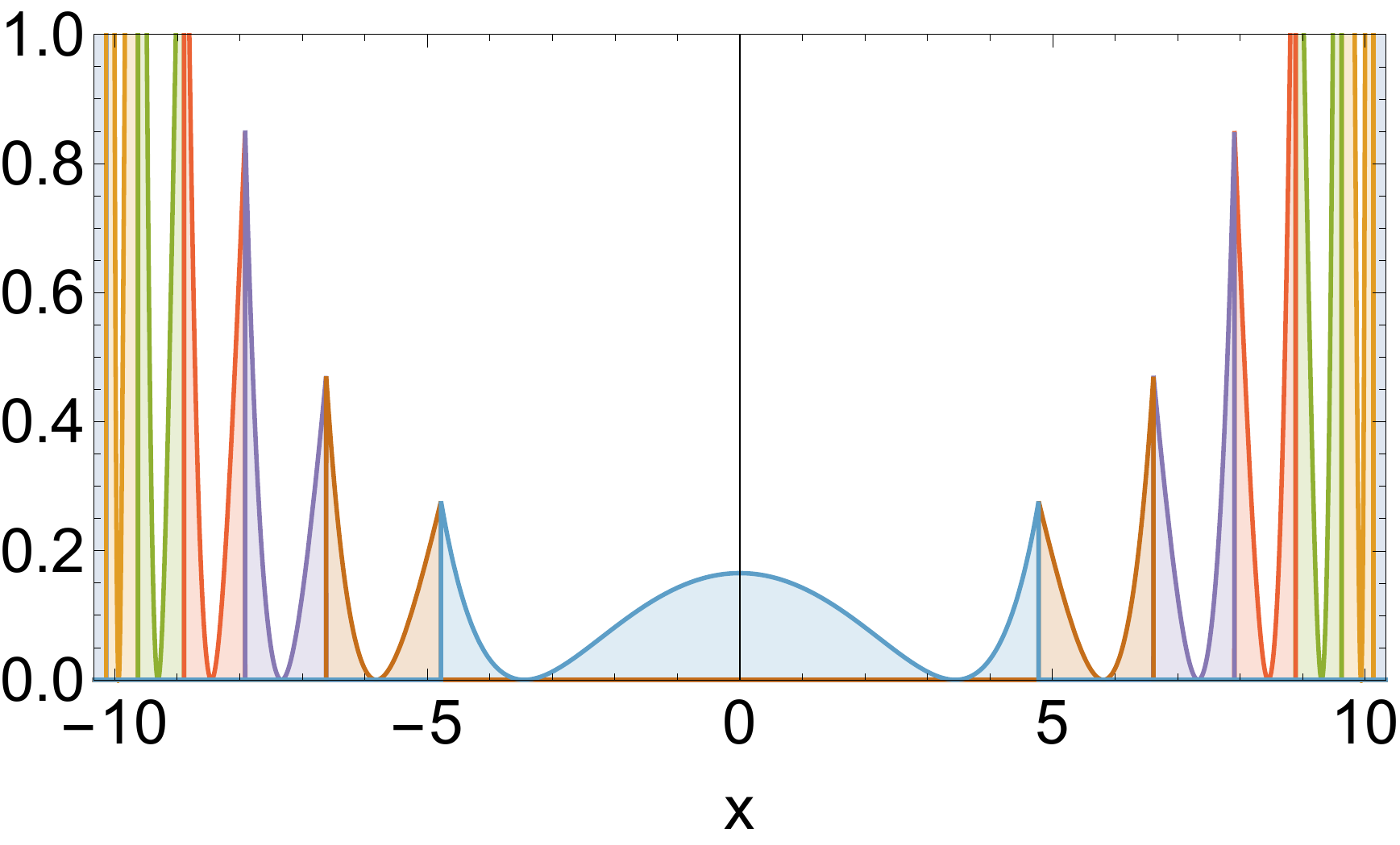}}
\hskip 0.5cm
\subfigure[$\quad |\phi_{k}(t,x)|,\quad k=0,1,\ldots,6$]{\includegraphics[width=0.45\textwidth,height=0.25\textwidth, angle =0]{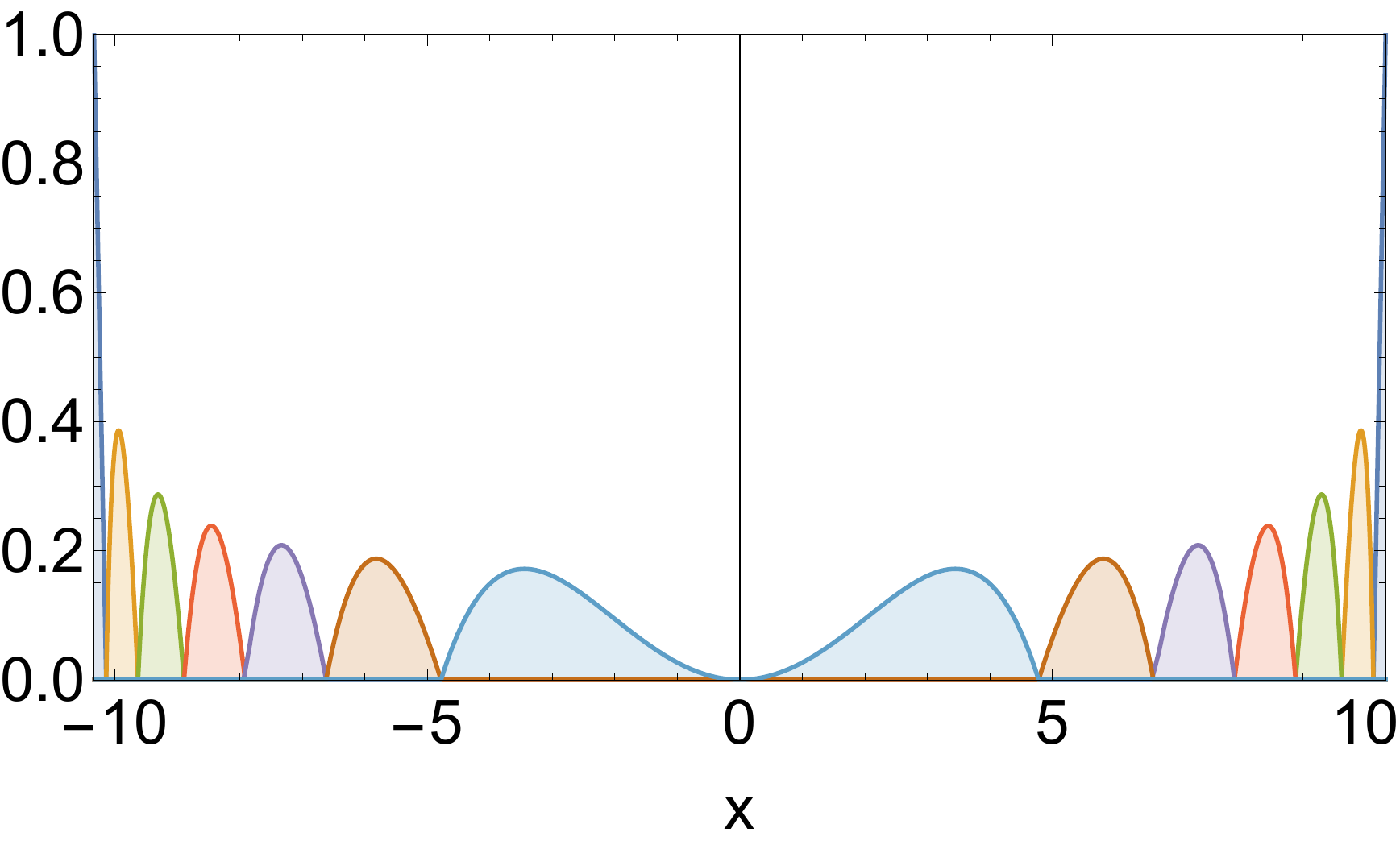}}
\caption{(a) Density of kinetic and gradient energy and (b) density of potential energy associated with the wave at $t=t_6$.}
\label{fig:densities}
\end{figure}

The energy  associated with the wave on the open segment $-t<x<t$ is given by
\be
E(t)=\sum_{k=0}^{N(t)}\Big(K_k(t)+U_k(t)\Big)
\ee
where $N(t)+1$ is a number of partial solutions at $t$ that form a shockwave and
\begin{align}
K_k(t)&:=\frac{1}{2}\int_{-t}^{t}dx\Big[(\partial_t\phi_k)^2+(\partial_x\phi_k)^2\Big]=\int_{{\rm supp}_{\phi_k}(t)}dx\left[\frac{x^2+t^2}{8}\Big(W_k'({\textstyle\frac{x^2-t^2}{4}})\Big)^2\right],\label{kin}\\
U_k(t)&:=\int_{-t}^{t}|\phi_k|dx=\int_{{\rm supp}_{\phi_k}(t)}dx\Big|W_k({\textstyle\frac{x^2-t^2}{4}})\Big|\label{pot}
\end{align}
describe contributions of individual partial solutions to the energy of the wave inside the lightcone. In Fig.\ref{fig:densities} we show integrands (densities) of \eqref{kin} and \eqref{pot} at $t=t_6$.

Both integrals presented above can be given by explicit functions. First we shall look at contributions to kinetic and gradient energy. We shall present separately the cases $k=0$ and $k=1,2,3,\ldots $ In the case $k=0$, we get $W'_0=1$. The expression $K_0(t)$ reads
\be
K_0(t)=2\int_{x_{\rm min}(t)}^{x_{\rm max}(t)}dx\frac{x^2+t^2}{8}=\frac{1}{4}\left[t^2x-\frac{x^3}{3}\right]_{x_{\rm min}(t)}^{x_{\rm max}(t)},\label{intK0}
\ee
where the factor ``$2$'' appears due to the spatial symmetry of the wave and where
\begin{eqnarray}
(x_{\rm min},x_{\rm max})=\left\{\begin{array}{lcr}
(0,t)&\quad{\rm for}\quad&0\le t\le t_0\\
(c_0(t), t)&\quad{\rm for}\quad&t\ge t_0
\end{array}\right.,\nonumber
\end{eqnarray}
with $t_0$ given by \ref{tk}.
The integral \eqref{intK0} reads
\be
K_0(t)=\left\{\begin{array}{lcr}
\frac{t^3}{3}&\quad{\rm for}\quad&0\le t\le t_0\\
\frac{t^3}{3}-\frac{1}{3}(t^2-a_0)c_0(t)&\quad{\rm for}\quad&t\ge t_0
\end{array}\right..\label{K0}
\ee
Expression $K_0(t)$ in the limit of large $t$ has linear behavior
\be
K_0(t)\approx a_0t+{\cal O}(t^{-3}).\label{assymptot0}
\ee

Energies $K_k(t)$ with $k=1,2,\ldots$ are given by expression
\begin{align}
K_k(t)&=2\int_{x_{\rm min}(t)}^{x_{\rm max}(t)}dx\frac{x^2+t^2}{8}\left(1+\frac{4b_k}{x^2-t^2}\right)^2\nonumber
\end{align}
where
\begin{eqnarray}
(x_{\rm min},x_{\rm max})=\left\{\begin{array}{lcr}
(0, c_{k-1}(t))&\quad{\rm for}\quad&t_{k-1}\le t\le t_k \\
(c_{k}(t), c_{k-1}(t))&\quad{\rm for}\quad&t\ge  t_k
\end{array}\right.\nonumber
\end{eqnarray}
and where $t_k$ and $c_k$ are given by \eqref{tk} and \eqref{ck}.  The kinetic and gradient part of the energy reads
\be
K_k(t)=\left\{\begin{array}{lcr}
K_k^{(-)}(t)=P_k(t)&\quad{\rm for}\quad& t_{k-1}\le t\le t_k\\
K_k^{(+)}(t)=P_k(t)+Q_k(t)&\quad{\rm for}\quad& t\ge  t_k
\end{array}\right.\label{Kk}
\ee
where we have defined symbols
\begin{align}
P_k(t)&:=\left[\frac{b_k^2}{a_{k-1}}+2b_k-\frac{a_{k-1}}{3}+\frac{t^2}{3}\right]c_{k-1}(t)-4b_k t\, {\rm ArcTanh\,}\left(\frac{c_{k-1}(t)}{t}\right)\nonumber,\\
Q_k(t)&:=-\left[\frac{b_k^2}{a_{k-1}}+2b_k-\frac{a_{k}}{3}+\frac{t^2}{3}\right]c_k(t)-4b_k t\, {\rm ArcTanh\,}\left(\frac{c_k(t)}{t}\right).\nonumber
\end{align}
Expression $K_k(t)$ has the following asymptotic behavior for $t\rightarrow\infty$
\be
K_k(t)\approx \left[\frac{b_k^2}{a_{k-1}}-\frac{b_k^2}{a_k}+a_{k-1}-a_k\right]t+{\cal O}(t^{-3}).\label{assymptotk}
\ee

\begin{figure}[h!]
\centering
\subfigure[$\quad K_k(t)$]{\includegraphics[width=0.45\textwidth,height=0.3\textwidth, angle =0]{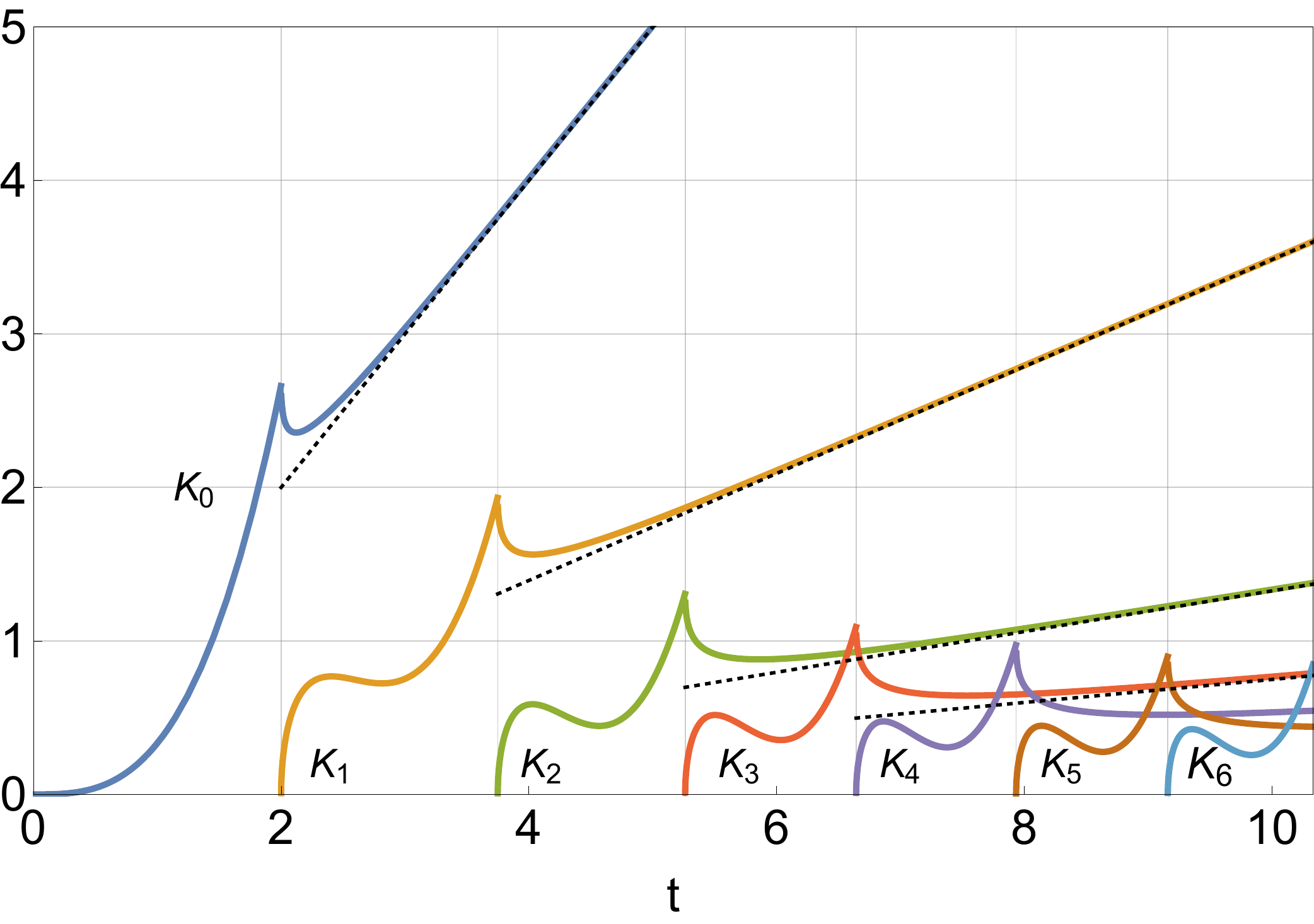}}
\hskip 0.5cm
\subfigure[$\quad U_k(t)$]{\includegraphics[width=0.45\textwidth,height=0.3\textwidth, angle =0]{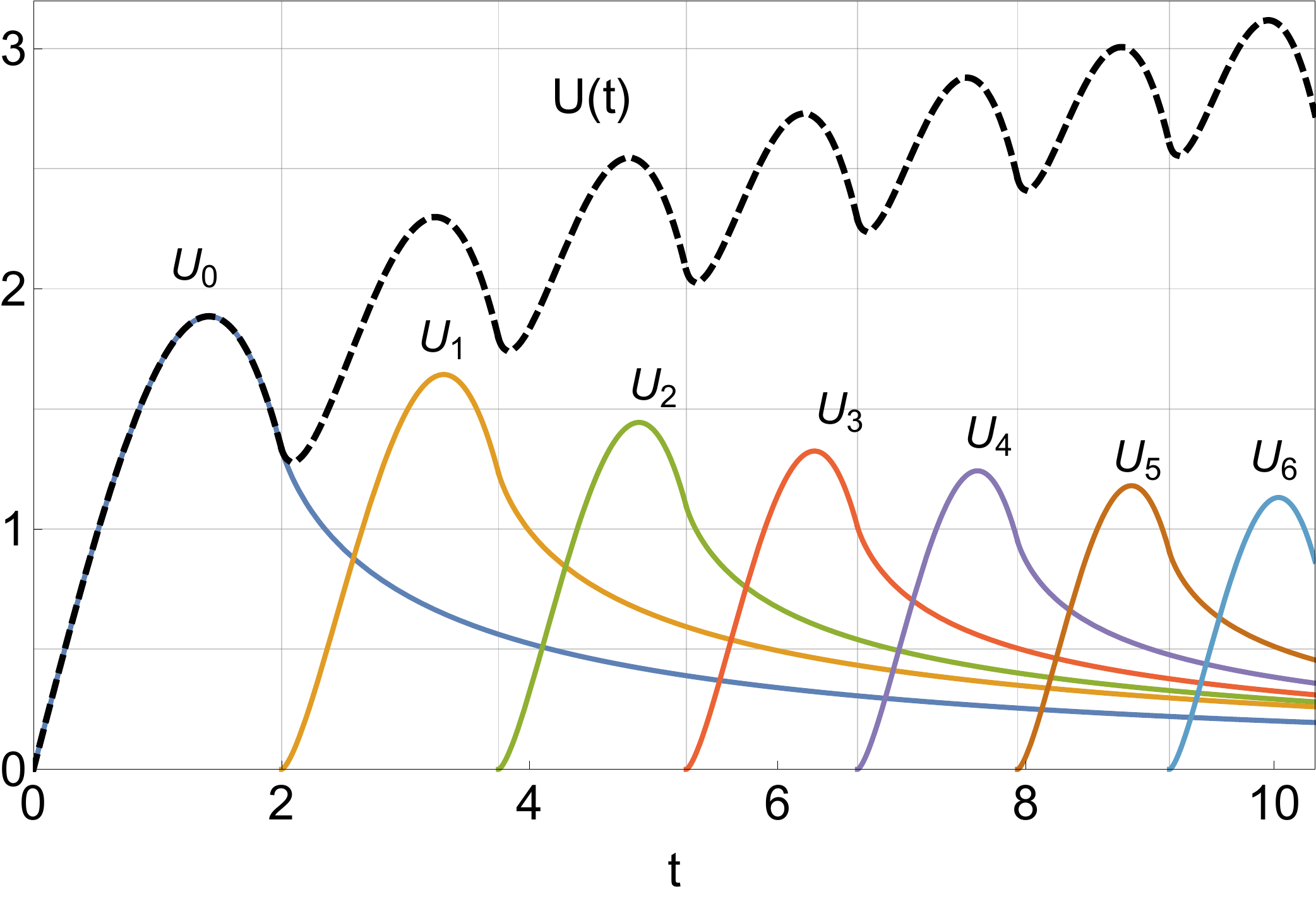}}
\caption{(a) Kinetic and gradient energies of partial solutions. (b) Potential energies of these solutions.}
\label{fig:integrals}
\end{figure}

In Fig.\ref{fig:integrals}(a) we show first seven contributions $K_k(t)$ associated with derivative terms in the energy density. The vertical grid lines correspond with instants of time $t_k=2\sqrt{a_k}$ at which new partial solutions appear. The abrupt decreasing of the energy $K_k(t)$ which manifests in the presence of spikes is caused by the appearance of new partial solution $\phi_{k+1}$ and consequently by a quick decrease of the support size of  $\phi_k$. We have plotted also the first four straight lines \eqref{assymptot0} and \eqref{assymptotk} for $k=1,2,3$. They are marked by dashed lines. The picture in Fig.\ref{fig:integrals}(a) shows that the dominating contribution to the kinetic and gradient energy is associated with the partial solution $k=0$. Its energy grows as $a_0 t$ for large $t$.\footnote{This is exactly half of the total energy of the wave, see \eqref{totalenergy}. } Note that the partial solution  $k=0$ is in contact with the discontinuity at $x=\pm t$. This discontinuity can be seen as a reservoir of infinite energy. The energy is  transferred  first into $\phi_0(t,x)$ and then it is transferred to the other partial solutions $\phi_k(t,x)$. The central partial solutions carry less energy than partial solutions localized close to the light cone.

Now we shall look at the potential energy of partial solutions. The potential energy associated with partial solution $\phi_0(t,x)$ reads
\be
U_0(t)=\left\{\begin{array}{lcr}
2a_0 t-\frac{t^3}{3}&\quad{\rm for}\quad&0\le t\le t_0\\
2a_0 t-\frac{t^3}{3}+\frac{1}{3}(t^2-4a_0)^{3/2}&\quad{\rm for}\quad&t\ge t_0
\end{array}\right.\label{U0}
\ee
which has the following asymptotic  behavior for large $t$
\be
U_0(t)\approx\frac{2a_0^2}{t}+{\cal O}(t^{-3}).
\ee
Similarly we find potential energy of  partial solutions with $k=1,2,\ldots$ They are given by
\be
U_k(t)=\left\{\begin{array}{lcr}
U_k^{(-)}(t)=R_k(t)&\quad{\rm for}\quad& t_{k-1}\le t\le t_k\\
U_k^{(+)}(t)=R_k(t)+S_k(t)&\quad{\rm for}\quad& t\ge  t_k
\end{array}\right.,\label{Uk}
\ee
where
\begin{align}
R_k(t)&=\left[\frac{4}{3}a_k-4b_k-\frac{t^2}{3}\right]c_{k-1}(t)+4b_k t\,{\rm ArcTanh\,}\left(\frac{c_{k-1}(t)}{t}\right),\nonumber\\
S_k(t)&=-\left[\frac{4}{3}a_{k-1}-4b_k-\frac{t^2}{3}\right]c_k(t)-4b_k t\,{\rm ArcTanh\,}\left(\frac{c_k(t)}{t}\right).\nonumber
\end{align}
We have eliminated logarithmic terms in $R_k$ and $S_k$ using the second one of relations  \eqref{rekurencja}.
The last formula implies that for large $t$ expression $U_k(t)$ behaves as
\be
U_k(t)\approx 2\left[(a_{k}-a_{k-1})(a_{k}+a_{k-1}-2b_k)\right] \frac{1}{t}+{\cal O}(t^{-2})
\ee
It shows that potential energy of each partial solution decreases as $t^{-1}$ for large $t$.

The energy of a single partial solution is given by a sum of expressions \eqref{K0} and \eqref{U0}
\be
E_0(t)=\left\{\begin{array}{lcr}
E_0^{(-)}(t)=X_0(t)&\quad{\rm for}\quad& 0\le t\le t_0,\\
E_0^{(+)}(t)=X_0(t)+Y_0(t)&\quad{\rm for}\quad& t\ge  t_0,
\end{array}\right.\label{E0}
\ee
where
\be
X_0(t):=2a_0t,\qquad Y_0(t):=-a_0c_0(t).
\ee
Similarly, for $k=1,2,\ldots$ the energy of partial solutions is obtained adding \eqref{Kk} and \eqref{Uk}. It reads
\be
E_k(t)=\left\{\begin{array}{lcr}
E_k^{(-)}(t)=X_k(t)&\quad{\rm for}\quad& t_{k-1}\le t\le t_k,\\
E_k^{(+)}(t)=X_k(t)+Y_k(t)&\quad{\rm for}\quad& t\ge  t_k,
\end{array}\right.\label{Ek}
\ee
where
\begin{align}
X_k(t)&:=P_k(t)+R_k(t)=\left[\frac{b_k^2}{a_{k-1}}-2b_k+a_{k-1}\right]c_{k-1}(t).\\
Y_k(t)&:=Q_k(t)+S_k(t)=-\left[\frac{b_k^2}{a_{k}}-2b_k+a_{k}\right]c_{k}(t).
\end{align}

\begin{figure}[h!]
\centering
{\includegraphics[width=0.6\textwidth,height=0.45\textwidth, angle =0]{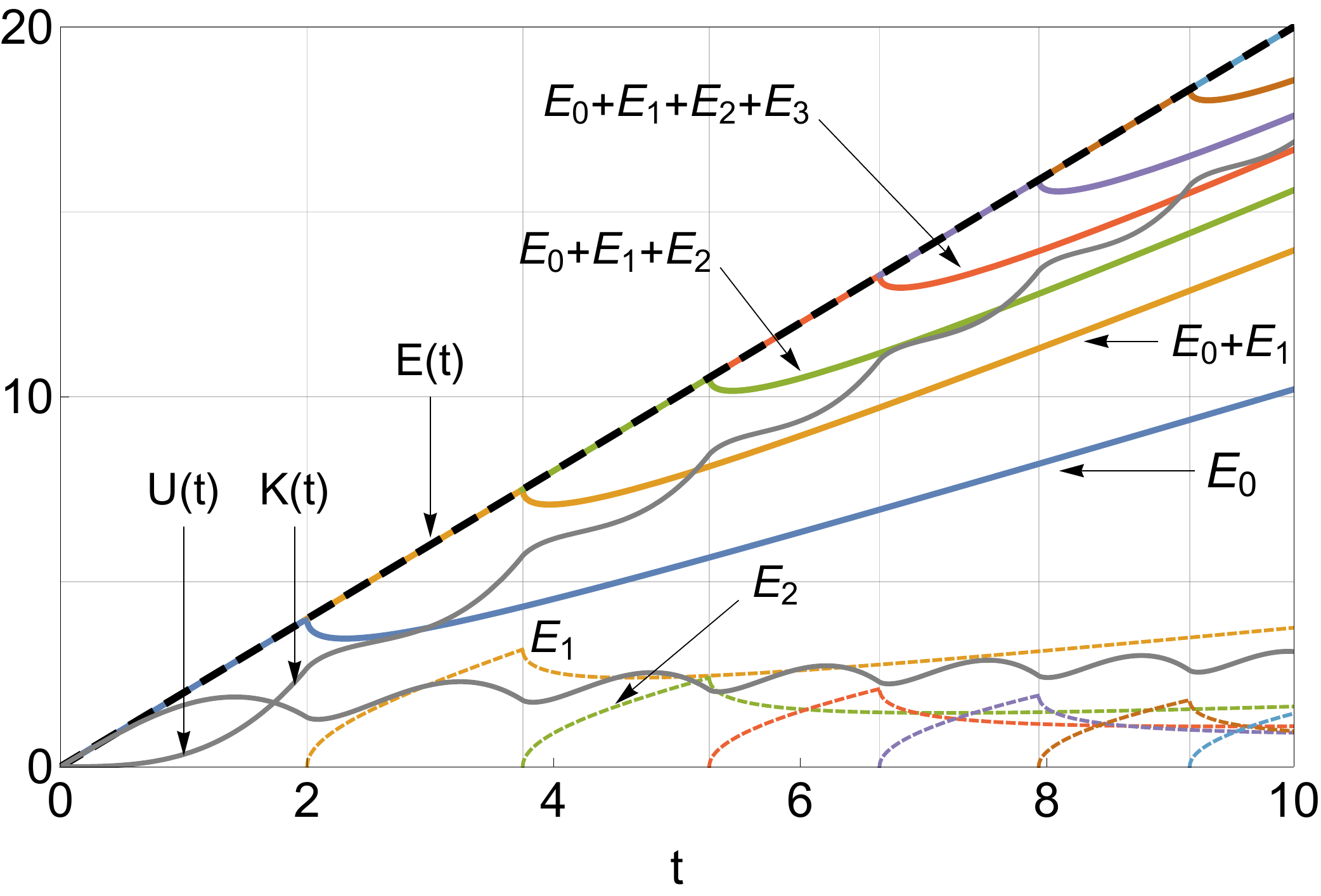}}
\caption{Total energies of partial solutions $E_k(t)$ and total energy $E(t)=K(t)+U(t)$ of the wave.  The dashed line represents total energy of the wave.}
\label{fig:integrals2}
\end{figure}

In Fig.\ref{fig:integrals2}  we plot energies \eqref{E0} and \eqref{Ek} of partial solutions together with individual contributions from kinetic-gradient and potential part. It is quite notable that the total energy of the wave grows exactly linearly with time.
In order to show this we consider a total energy of the wave at $t$ belonging to the interval $ t_{k-1}\le t\le t_k$. Since the wave consists on first $k+1$ partial solutions then it has the energy
\be
E(t)=\sum_{i=0}^kE_i(t)=\sum_{i=0}^{k-1}E^{(+)}_i(t)+E_k^{(-)}(t)\label{totalenergy}
\ee
where $E^{(+)}_i(t)=X_i(t)+Y_i(t)$ and $E_k^{(-)}(t)=X_k(t)$. Expression \eqref{totalenergy} can be cast in the form
\begin{align}
E(t)=X_0(t)+\sum_{i=1}^{k}[Y_{i-1}(t)+X_i(t)].\nonumber
\end{align}
Making use of the relation $b_1=2a_0$ and the first one of relations \eqref{rekurencja} we get
\begin{align}
X_1(t)+Y_0(t)&=\left[\frac{b_1^2}{a_0}-2b_1\right]c_0(t)=0,\nonumber\\
X_i(t)+Y_{i-1}(t)&=\left[\frac{b^2_i-b^2_{i-1}}{a_{i-1}}-2b_{i-1}-2b_i\right]c_{i-1}(t)=0, \qquad i=1,2,\ldots k.\nonumber
\end{align}
It shows that the energy of the wave inside the light cone is given by expression $X_0(t)$
\be
E(t)=2a_0 t.\label{totalenergy2}
\ee
This result shows that energy of the wave inside the light cone grows linearly with time.

\section{Solutions}
In this section we go back to regular initial data with delta-like profile of time derivative of the field. First we present numerical solutions of shock-like waves and then give analytical solution for initial phase of formation of the wave. Finally, we compare some numerical simulations containing formation of shock-like waves with results of scattering of compact oscillons.

\subsection{Numerical results for triangular and Gaussian initial data}



Here we shall present results of numerical evolution of initial data
\be
\phi(t,x)|_{t=0}=0,\qquad \partial_t\phi(t,x)|_{t=0}=a\,\delta_{\epsilon}(x)\label{inireg}
\ee
where $\delta_{\epsilon}(x)$ is given by both functions the Gaussian one \eqref{gauss} and the triangular one \eqref{triangle}. We study how numerical solution changes in dependence on the value of parameter $\epsilon$. The results are presented on spacetime diagrams where the value of the field marked by gradient color. We also plot  the energy density of the system
\begin{equation}
{\cal H}=\frac{1}{2}(\partial_t\phi)^2+\frac{1}{2}(\partial_x\phi)^2+|\phi|.
\end{equation}



\begin{figure}[]
\centering
\subfigure[$\quad \epsilon=10^{-6}$]{\includegraphics[width=0.44\textwidth,height=0.25\textwidth, angle =0]{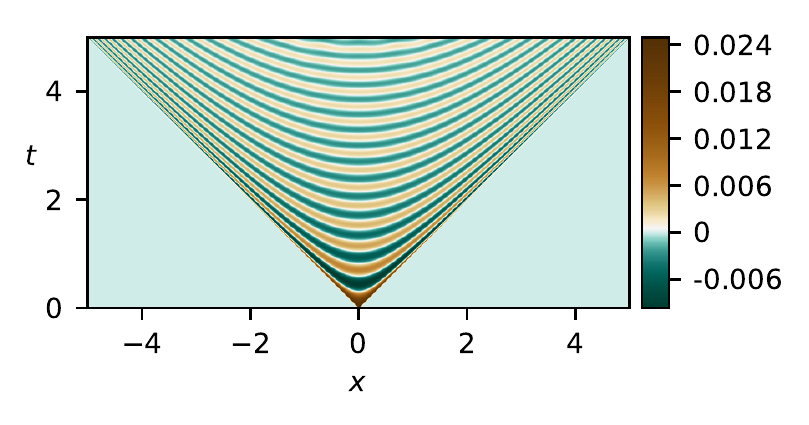}}
\subfigure[$\quad \epsilon=10^{-6}$]{\includegraphics[width=0.44\textwidth,height=0.25\textwidth, angle =0]{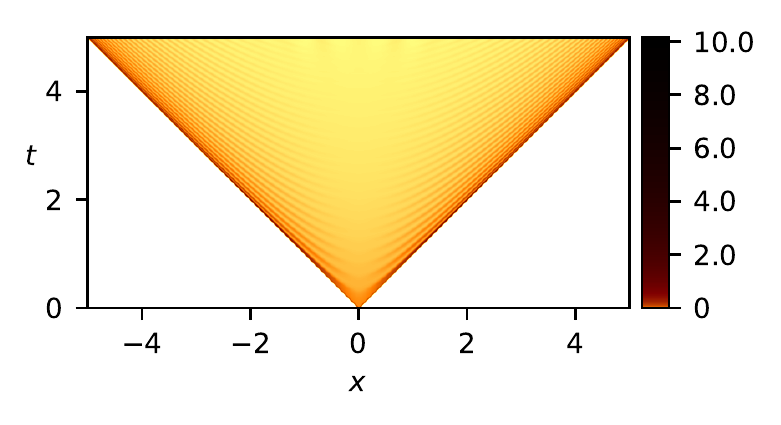}}
\subfigure[$\quad \epsilon=2 \cdot 10^{-5}$]{\includegraphics[width=0.44\textwidth,height=0.25\textwidth, angle =0]{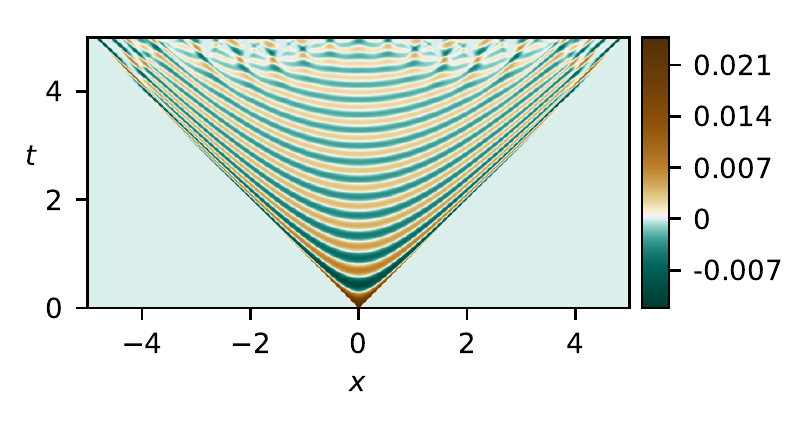}}
\subfigure[$\quad \epsilon=2 \cdot 10^{-5}$]{\includegraphics[width=0.44\textwidth,height=0.25\textwidth, angle =0]{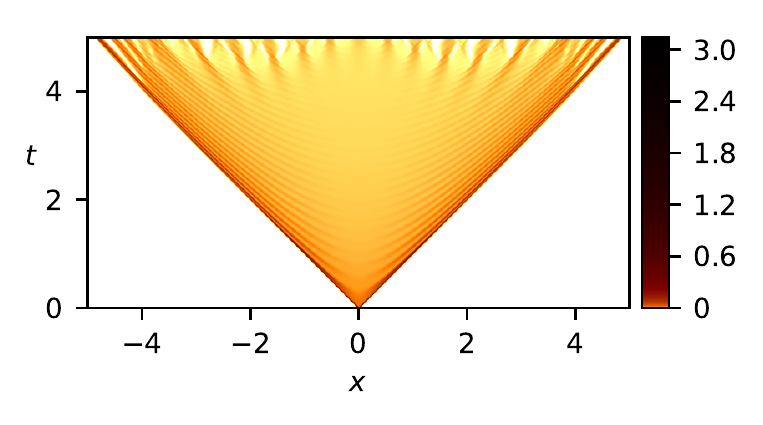}}
\subfigure[$\quad \epsilon=4  \cdot10^{-5}$]{\includegraphics[width=0.44\textwidth,height=0.25\textwidth, angle =0]{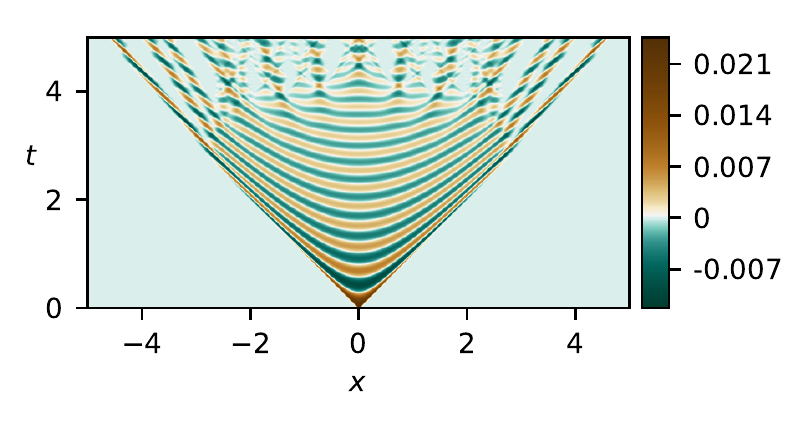}}
\subfigure[$\quad \epsilon=4  \cdot10^{-5}$]{\includegraphics[width=0.44\textwidth,height=0.25\textwidth, angle =0]{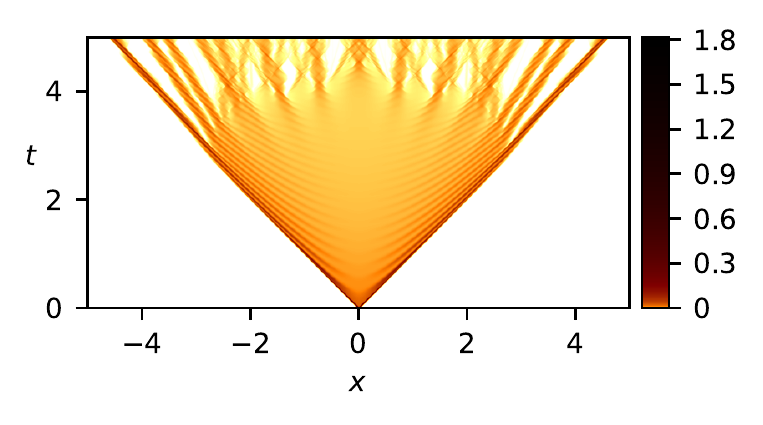}}
\subfigure[$\quad \epsilon= 5 \cdot10^{-5}$]{\includegraphics[width=0.44\textwidth,height=0.25\textwidth, angle =0]{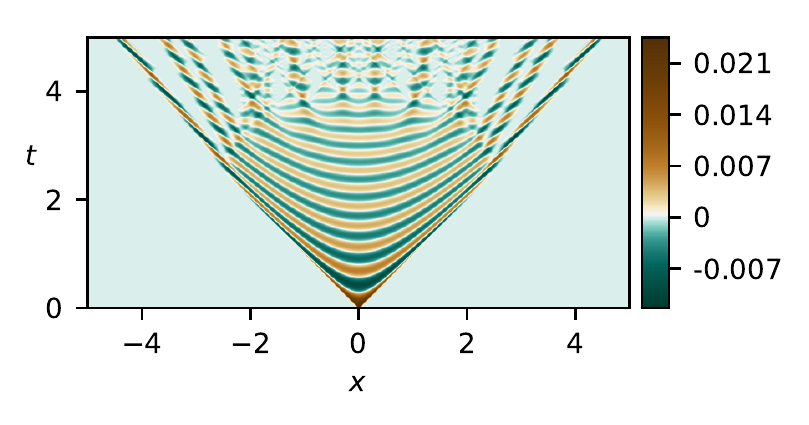}}
\subfigure[$\quad \epsilon=5  \cdot10^{-5}$]{\includegraphics[width=0.44\textwidth,height=0.25\textwidth, angle =0]{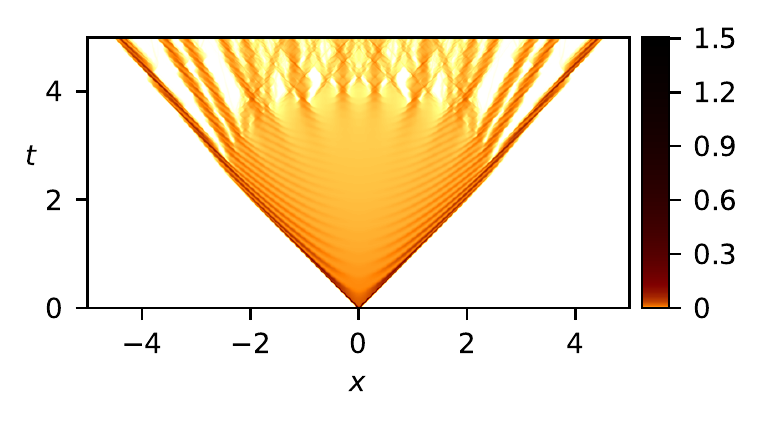}}
\caption{The case of triangular form of $\delta_{\epsilon}$. The decay of the shock wave in function of $\epsilon$ for $a=0.05$. Field $\phi(t,x)$ (left) and its energy density (right).}
\label{fig:t1}
\end{figure}

In Fig.\ref{fig:t1} we show evolution of the signum-Gordon field for delta-like initial profile of $\partial_t\phi$. We consider $\epsilon=\{10^{-6}, 2\cdot10^{-5}, 4\cdot 10^{-5}, 5\cdot 10^{-5}\}$. For  $\epsilon=10^{-6}$ the numerical solution looks very similar to exact shock wave. After increasing $\epsilon$ by factor $20$ we see that the wave breaks down for $t > 4$. Looking at the energy density in Fig.\ref{fig:t1}(d) we can see an initial phase of formation of jets. This behavior is even more visible for $\epsilon=4\cdot 10^{-5}$ and $\epsilon=5\cdot 10^{-5}$. Looking in more detail at these jets we see that they contain structures which are very similar to oscillons. Some of this oscillons interact with others what leads to quite complex structures. The region of existence of a shock-like wave configuration of the signum-Gordon field shrinks when $\epsilon$ decreases.

\begin{figure}[h!]
\centering
\subfigure[$\quad \phi(t,x)$]{\includegraphics[width=0.44\textwidth,height=0.3\textwidth, angle =0]{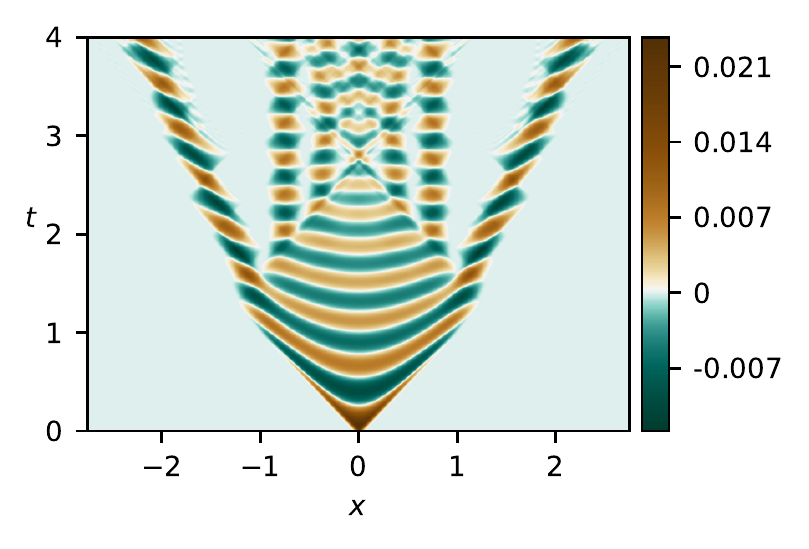}}
\subfigure[$\quad {\cal H}(t,x)$]{\includegraphics[width=0.44\textwidth,height=0.3\textwidth, angle =0]{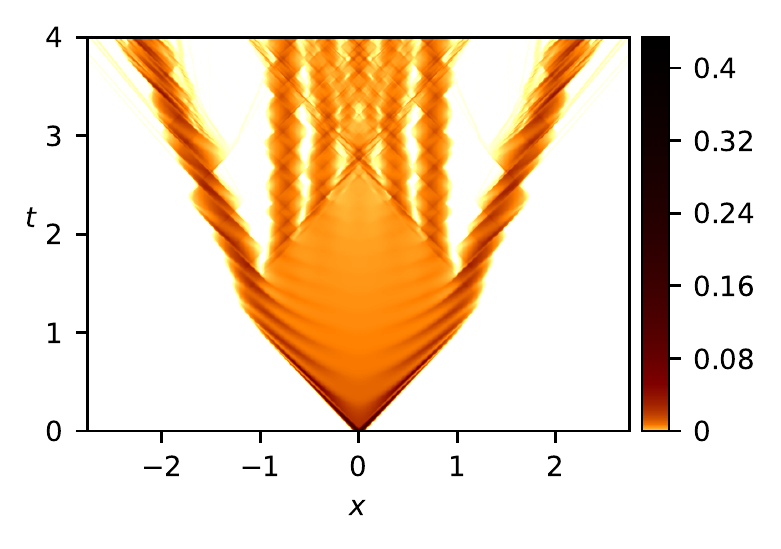}}
\caption{The case of triangular form of $\delta_{\epsilon}$: $a=0.05$ and $\epsilon=2\times 10^{-4}$.  (a) Field $\phi$ and (b) its energy density.}
\label{fig:t2}
\end{figure}

We have also looked at higher values of $\epsilon$. For $\epsilon=2\cdot10^{-4}$, see Fig.\ref{fig:t2}(a,b), the region of spacetime being a support of the wave has a diamond-shaped form.  At left and right edge of the diamond emerge two oscillons. They move in opposite directions with relatively high speeds. In the central part of the diagram we see formation of certain number of slow oscillons. They interact with each other by scattering and by emission and absorption of radiation.

\begin{figure}[h!]
\centering
\subfigure[$\quad \phi(t,x)$]{\includegraphics[width=0.44\textwidth,height=0.3\textwidth, angle =0]{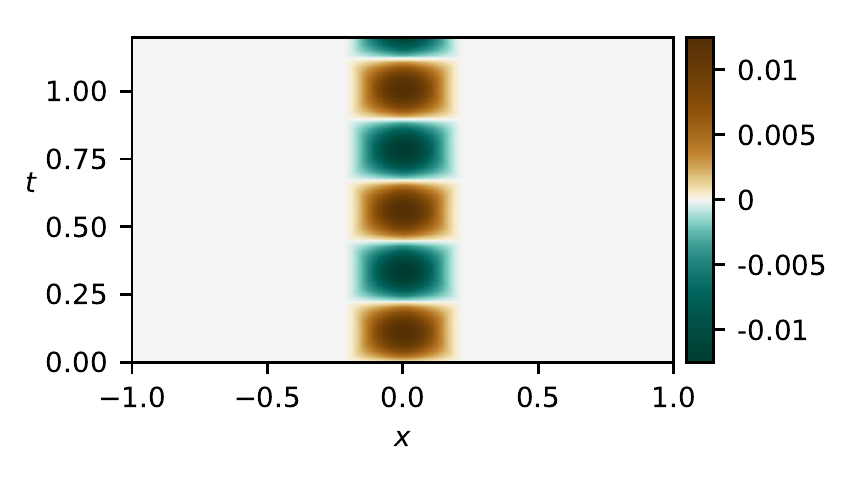}}
\subfigure[$\quad {\cal H}(t,x)$]{\includegraphics[width=0.44\textwidth,height=0.3\textwidth, angle =0]{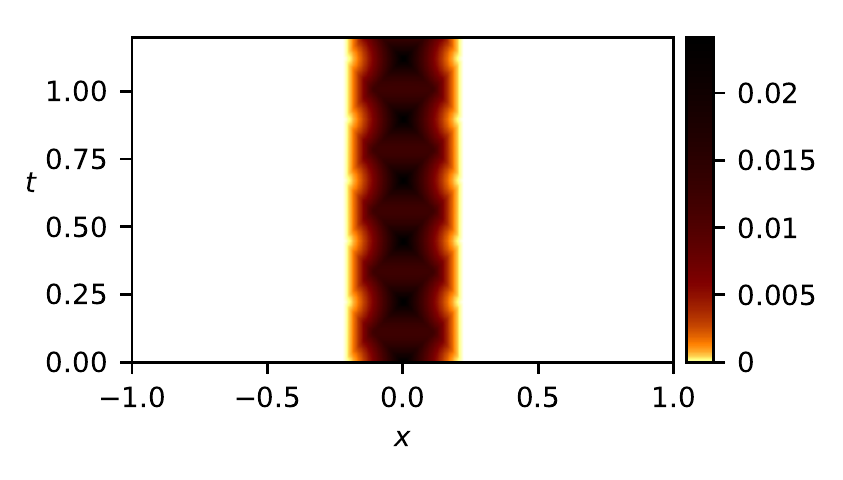}}
\caption{The case of triangular form of $\delta_{\epsilon}$: $a=0.05$ and $\epsilon=0.0039788\approx4\times 10^{-3}$. In this case the numerical solution corresponds with the exact oscillon.}
\label{fig:t3}
\end{figure}

Our choice of triangular shape of $\delta_{\epsilon}$ function allows to obtain an exact oscillon. In the case of the simplest exact oscillon the initial profile of velocity is an isosceles triangle with sides that form angle $\alpha=\pi/4$ with axis $x$. Hence, the ratio of its height $a\delta_{\epsilon}(0)=\frac{a}{2\sqrt{\pi \epsilon}}$ by a half of size of its base $2\sqrt{\pi\epsilon}$ must be equal to unity, i.e. $\frac{a}{4\pi\epsilon}=1$. It means that $\epsilon=\frac{a}{4\pi}$ is a special value that gives  the simplest exact oscillon with the support size $2\sqrt{a}$. Taking  $a=0.05$ one gets $\epsilon=\frac{1}{80\pi}\approx0.0039788\approx4\cdot 10^{-3}$. The numerical solution corresponding with this case is presented in Fig.\ref{fig:t3}.

\begin{figure}[h!]
\centering
\subfigure[$\quad \phi(t,x)$]{\includegraphics[width=0.44\textwidth,height=0.3\textwidth, angle =0]{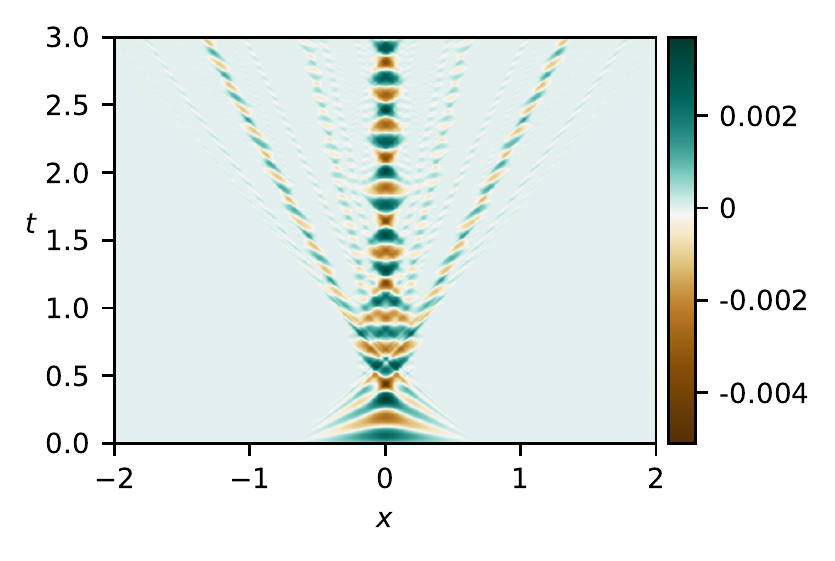}}
\subfigure[$\quad {\cal H}(t,x)$]{\includegraphics[width=0.44\textwidth,height=0.3\textwidth, angle =0]{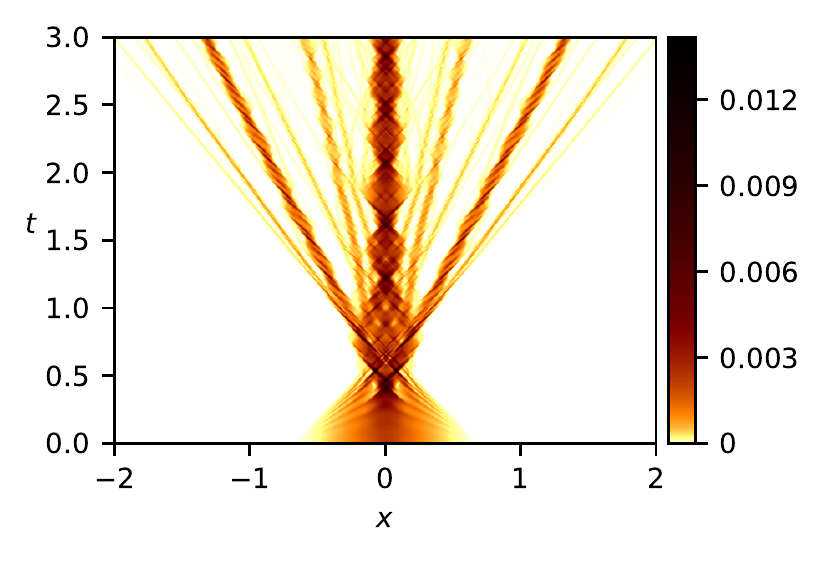}}
\caption{The case of triangular form of $\delta_{\epsilon}$: $a=0.05$ and $\epsilon=4\times 10^{-2}$.}
\label{fig:t4}
\end{figure}
We have also checked the evolution of the signum-Gordon field for higher values of $\epsilon$. A solution obtained for $\epsilon=4\cdot 10^{-2}$ is sketched in  Fig.\ref{fig:t4}. The numerical solution represent collision of two segments of self-dual solutions which results in appearance of radiation that consists on many oscillons.


\begin{figure}[h!]
\centering
\subfigure[$\quad \phi(t,x)$]{\includegraphics[width=0.44\textwidth,height=0.3\textwidth, angle =0]{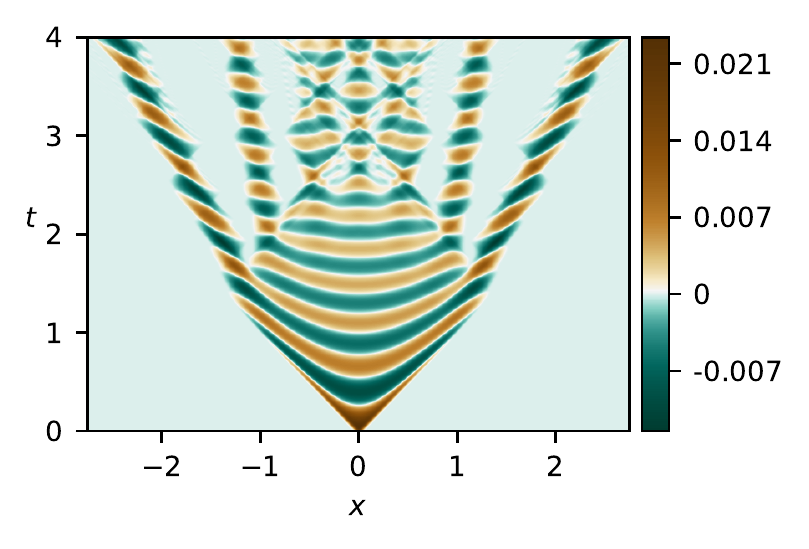}}
\subfigure[$\quad {\cal H}(t,x)$]{\includegraphics[width=0.44\textwidth,height=0.3\textwidth, angle =0]{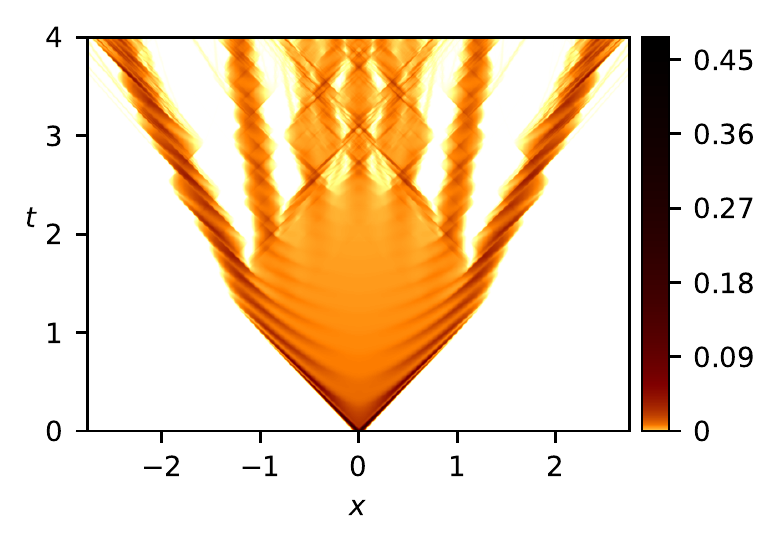}}
\caption{The case of Gaussian form of $\delta_{\epsilon}$: $a=0.05$ and $\epsilon=2\times 10^{-4}$. (a) Field $\phi$ and (b) its energy density.}
\label{fig:t5}
\end{figure}

\begin{figure}[h!]
\centering
\subfigure[$\quad \phi(t,x)$]{\includegraphics[width=0.44\textwidth,height=0.3\textwidth, angle =0]{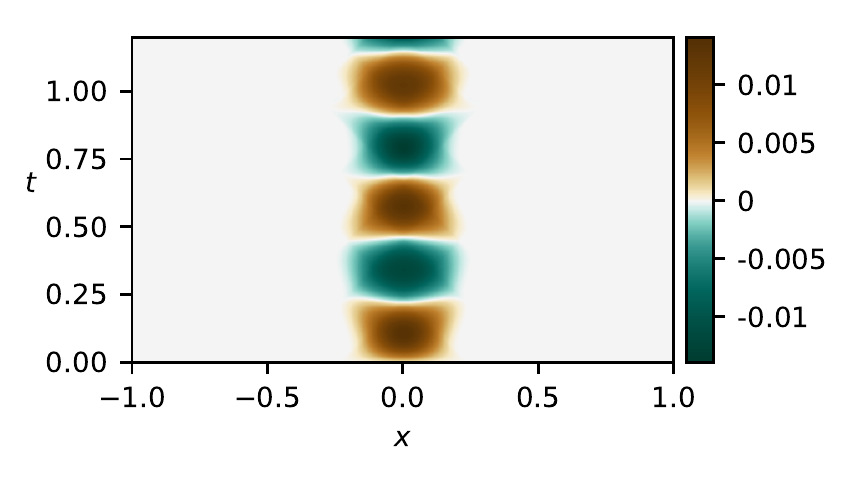}}
\subfigure[$\quad {\cal H}(t,x)$]{\includegraphics[width=0.44\textwidth,height=0.3\textwidth, angle =0]{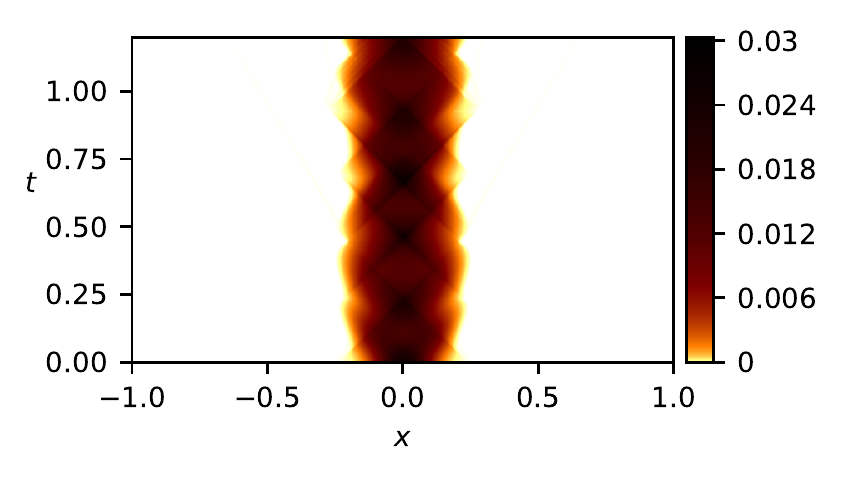}}
\caption{The case of Gaussian form of $\delta_{\epsilon}$: $a=0.05$ and $\epsilon=4\times 10^{-3}$.}
\label{fig:t6}
\end{figure}

\begin{figure}[h!]
\centering
\subfigure[$\quad \phi(t,x)$]{\includegraphics[width=0.44\textwidth,height=0.3\textwidth, angle =0]{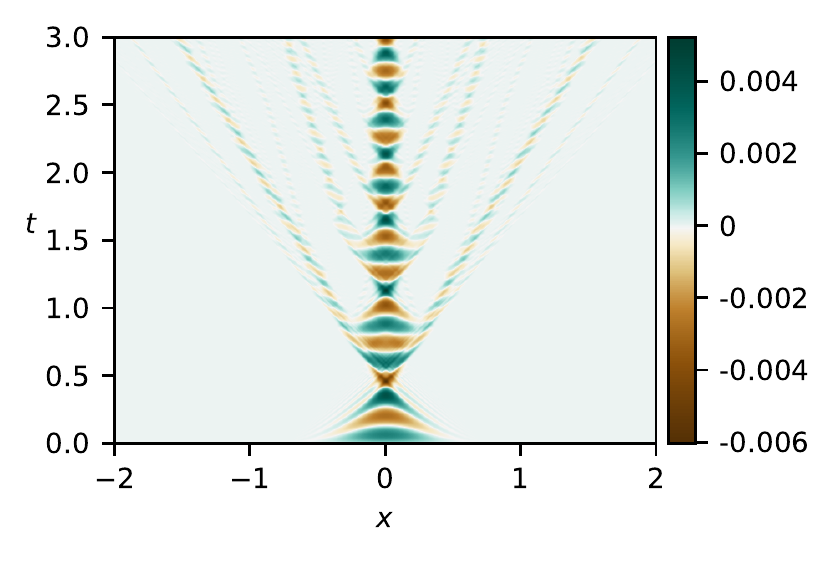}}
\subfigure[$\quad {\cal H}(t,x)$]{\includegraphics[width=0.44\textwidth,height=0.3\textwidth, angle =0]{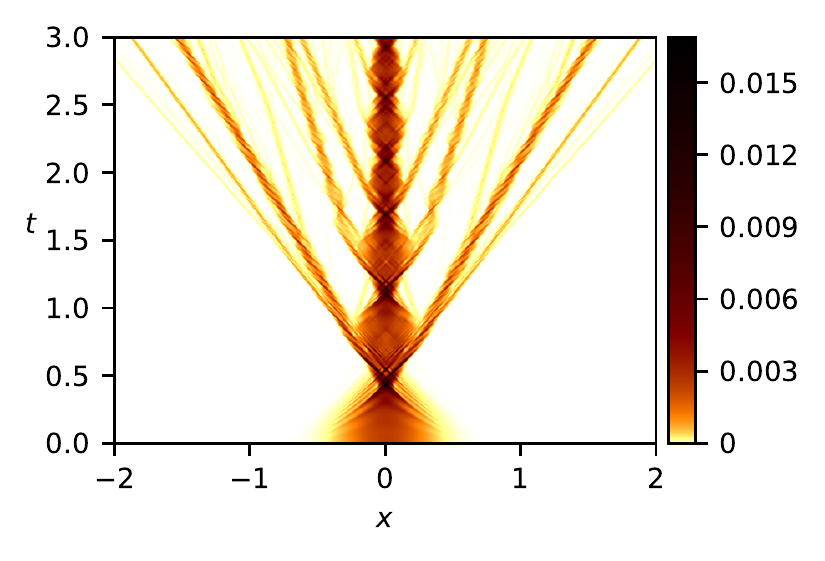}}
\caption{The case of Gaussian form of $\delta_{\epsilon}$: $a=0.05$ and $\epsilon=4\times 10^{-2}$.}
\label{fig:t7}
\end{figure}

In order to check to which extend our results depend on the choice of the form of $\delta_{\epsilon}(x)$ we also performed simulations taking its Gaussian version. The results are shown in  Figs.\ref{fig:t5} --\ref{fig:t7}. Comparing these figures with the previous ones we do not see much difference between triangular and Gaussian case.  Figures Fig.\ref{fig:t2} and Fig.\ref{fig:t5} are very similar. More differences appear when comparing the Gaussian case for $\epsilon=\frac{1}{80\pi}$ with the triangular case suitable for exact oscillon. In contrary to Fig.\ref{fig:t3} the solution shown in Fig.\ref{fig:t6} has a border which is not a straight line. This is pretty clear from the picture of the energy density shown in Fig.\ref{fig:t7}(b).  A counterpart of the exact oscillon obtained in this case is a solution whose support shrinks and expands periodically. Similarly in the case $\epsilon=4\cdot 10^{-2}$ a region with high value of the energy density shrinks and expands with certain regularity, see Fig.\ref{fig:t7}. The maximum size of the regions regions with high value of the energy density decreases with time because of emission of oscillons.

\subsection{Exact solution in an initial phase of evolution}

In this section we shall study some analytical expressions that describe a shock-like wave solution.  Having in mind the results of our numerical simulations we do not expect  to obtain exact solution for arbitrarily long times. On the other hand, the initial triangular data are simple enough to get some analytical results describing a wave in its initial stage of evolution. It would be interesting to compare such solution with their counterpart that form an exact shock wave.

 The triangular initial data \eqref{inireg} with $\epsilon\equiv\frac{\varepsilon^2}{4\pi}$  are given on a segment $x\in[-\varepsilon, \varepsilon] $ where the triangular shape is given by expression
 \[
 \delta_{\varepsilon}(x)=\frac{1}{\varepsilon^2}\Big[(\varepsilon+x)\theta(\varepsilon+x)\theta(-x)+(\varepsilon-x)\theta(\varepsilon-x)\theta(x)\Big].
 \]
The energy of such initial field configuration equals to \eqref{E2} or $E=\frac{a^2}{3\varepsilon}$. This energy  is finite and conserved during the evolution.
The initial data are symmetric under spatial reflection and so is the solution. This solution consists on partial solutions that have the general form given by expression \eqref{rozw}. It is known that initial data for the signum-Gordon field determine the number and the size of supports of partial solutions. Restriction to  triangular initial data still leaves a freedom of a choice of parameters $a$ and $\varepsilon$.  These parameters determine an inclination of the sides of a triangle.  According to Ref.\cite{ss} an inclination of the velocity profile determines whether the support of the solution spreads out, remain unchanged (the exact oscillon) or even shrinks.  Since we are interested in solutions which are similar to exact shock waves  then we will study a triangular shape with height $a\delta_{\varepsilon}(0)=\frac{a}{\varepsilon}>\varepsilon$.  An  example of such a solution with  $a=0.05$ and $\epsilon=2\times 10^{-4}$ (thus $\frac{a}{\varepsilon^2}=\frac{a}{4\pi\epsilon}\approx19.89>1$) is shown in Fig.\ref{fig:t2}.
\begin{figure}[h!]
\centering
\subfigure[]{\includegraphics[width=0.44\textwidth,height=0.25\textwidth, angle =0]{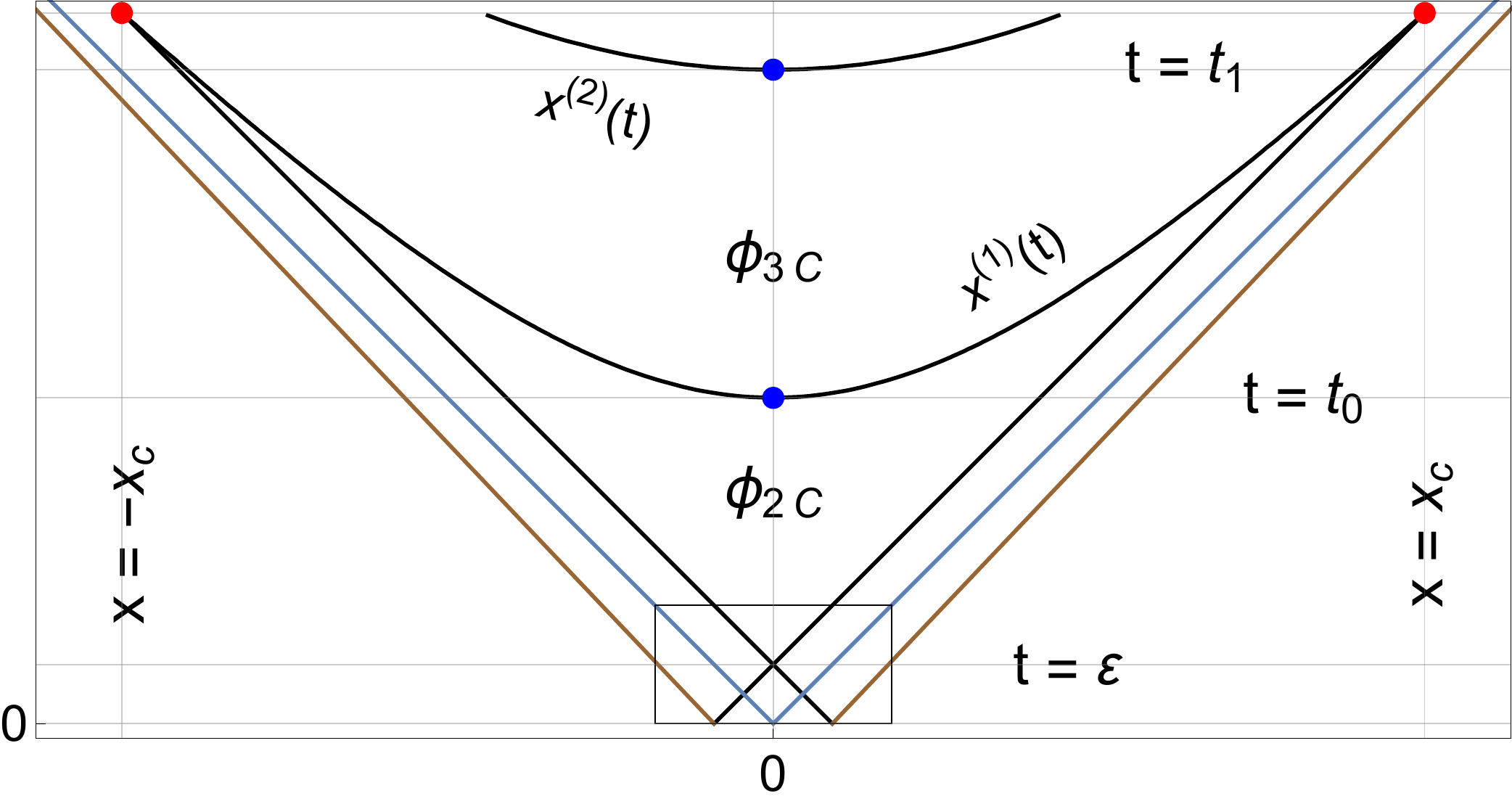}}
\hskip0.5cm
\subfigure[]{\includegraphics[width=0.44\textwidth,height=0.25\textwidth, angle =0]{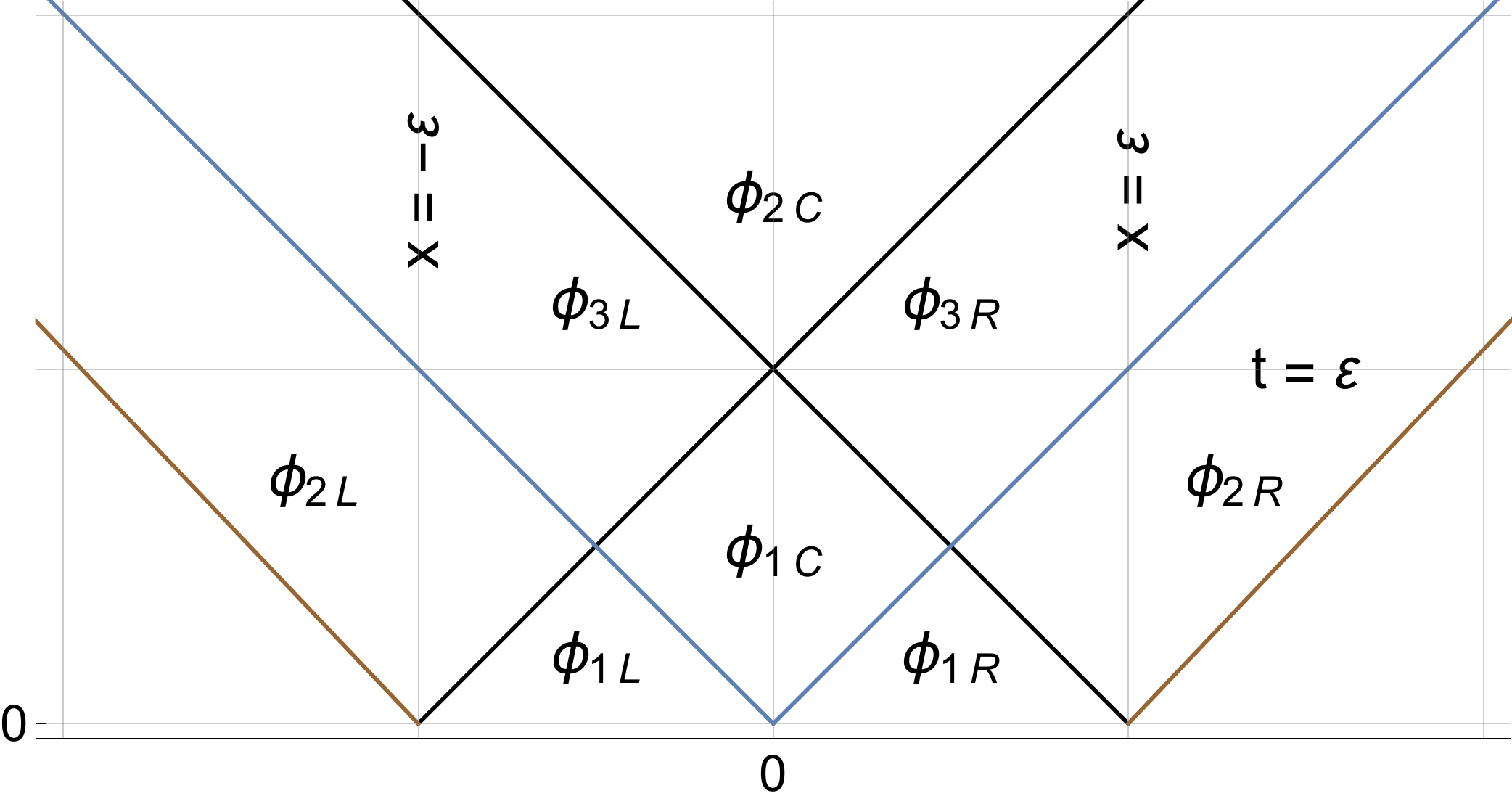}}
\caption{Partial solutions for triangular initial data: $a=0.05$ and $\epsilon=2\times 10^{-4}$ ($\varepsilon=0.05013$).}
\label{fig:diag}
\end{figure}
In Fig.\ref{fig:diag} we present domains of various partial solutions that compose the solution in its initial stage of evolution. Figure (b) shows blowup of the rectangular region at figure (a).  The partial solutions $\phi_{1L}(t,x)$ can be obtained directly  from initial data and it reads
\be
\phi_{1L}(t,x)=\frac{a}{\varepsilon^2}[x+\varepsilon]t-\frac{t^2}{2}.\label{phisgL1}
\ee
Its symmetric counterpart is given by $\phi_{1R}(t,x)=\phi_{1L}(t,-x)$.
These two solutions are restricted to the interior of past light cones of two events $(t,x)=(\frac{\varepsilon}{2},\pm\frac{\varepsilon}{2})$. Note that \eqref{phisgL1} and \eqref{phiweL1} differ only by expression $-\frac{t^2}{2}$.

Partial solution $\phi_{2L}(t,x)$  interpolate between $\phi_{1L}(t,x)$ and the vacuum solution $\phi=0$. It matches $\phi_{1L}(t,x)$ at the light cone $x=-\varepsilon+t$ and the vacuum at $x=-\varepsilon +v t$, where $v<0$. This solution reads
\be
\phi_{2L}(t,x)=\frac{(x+\varepsilon-v t)^2}{2(1-v^2)},\qquad v=-1+\frac{\varepsilon^2}{a}.\label{phi2Lreg}
\ee
Its symmetric counterpart is given by $\phi_{2R}(t,x)=\phi_{2L}(t,-x)$. Note that subluminal zero $x=-\varepsilon+vt$ crosses the lightcone $x=-t$ at $t_c=\frac{a}{\varepsilon}$. It results in compactness of a region of spacetime in which $\phi_{2L}$ holds. Note also that $\phi_{1L}$ and $\phi_{2L}$ (as well as $\phi_{1R}$ and $\phi_{2R}$) are given by expressions which are  identical with formulas that describe self-similar solutions discussed in \cite{ss}. The basic difference is that in the present case the supports  of self-similar partial solutions are restricted to  a region corresponding with exterior of the future light cone of the event $(t,x)=(0,0)$. The  region inside this light cone contains some new partial solutions.

In order to obtain the partial solution $\phi_{1C}(t,x)$ we assume its positivity (i.e. we choose a term ``$-\frac{t^2}{2}$'') and impose the following matching conditions $\phi_{1C}(t,-t)=\phi_{1L}(t,-t)$ and $\phi_{1C}(t,t)=\phi_{1R}(t,t)$. It gives
\be
\phi_{1C}(t,x)=\frac{1}{1+v}\left[\varepsilon t-\frac{x^2+t^2}{2}\right]-\frac{t^2}{2}.
\ee
This solution remains valid in a compact region delimited by intersection of the future light cone of the event $(0,0)$ and the past light cone of the event $(\varepsilon,0)$. The events $(\frac{\varepsilon}{2},\pm\frac{\varepsilon}{2})$ correspond with two points at the Minkowski diagram at which new partial solutions $\phi_{3L}$ and $\phi_{3R}$ emerge. Solution $\phi_{3L}(t,x)$ satisfies matching conditions $\phi_{3L}(t,-t)=\phi_{2L}(t,-t)$ and $\phi_{3L}(t,-\varepsilon+t)=\phi_{1C}(t,-\varepsilon+t)$. It takes the form
\begin{align}
 \phi_{3L}(t,x)=&\frac{1}{1+v}\left[\frac{\varepsilon}{2}(x+t)-\frac{1}{4}(x+t)^2\right]+\frac{(x-t)^2}{8}-\frac{t^2}{2}\nonumber\\&+\frac{1}{2(1-v^2)}\left[\varepsilon+(1+v)\frac{x-t}{2}\right]^2.
\end{align}

Fig.\ref{fig:diag} shows that in the limit $\varepsilon\rightarrow 0$  all the supports of partial solutions presented above shrink to points. Moreover,
in this limit the gradients of partial solutions became singular. It reflects emergence of discontinuities of the field at the light cone. As the result, there are no counterparts for these solutions in the set of partial solutions describing the exact shock wave.

On the other hand, partial solution $\phi_{2C}(t,x)$ is a counterpart of first partial solution $\phi_0(t,x)=\frac{1}{4}(x^2-t^2)+a_0$ that composes the exact shock wave. This solution matches $\phi_{3L}(t,x)$ at $x=\varepsilon-t$ and  $\phi_{3R}(t,x)$ at $x=-\varepsilon+t$. It has the form
\be
\phi_{2C}(t,x)=\frac{x^2-b(t)}{2(1-v)}\qquad{\rm where}\qquad b(t):=-vt^2+2\varepsilon t-2a.\label{phi2Creg}
\ee
Indeed, in the limit $\varepsilon\rightarrow 0$, ($v\rightarrow -1$) the solution $\phi_{2C}(t,x)$ tends to $\phi_0(t,x)$ with $a_0=\frac{1}{2}a$ (see Fig.\ref{fig:shock}(a) for meaning of $a_0$).  The coefficient $b(t)$ satisfies $b(t)\ge 0$  for $t\le t_0$  where
\be
 t_0=-\frac{\varepsilon}{v}\left[\sqrt{\frac{1-v}{1+v}}-1\right].
\ee
Thus $\phi_{2C}(t_0,0)=0$. Note that $t_0\rightarrow\sqrt{2a}=2\sqrt{a_0}$ in the limit $\varepsilon\rightarrow 0$. For $t>t_0$ the zero of $\phi_{2C}(t,x)$ splits into two zeros $x^{(1)}(t)=\pm\sqrt{b(t)}$ that move in opposite directions. Trajectories of zeros $x^{(1)}(t)$ delimit a domain of solution $\phi_{2C}(t,x)$. In the case of exact shock wave ($\varepsilon=0$) this region is restricted from below by light cone $x=\pm t$ and from above by a hyperbola $x(t)=\pm\sqrt{t^2-2a}$. This situation changes qualitatively for solution \eqref{phi2Creg} which, in contrary to its counterpart, is delimited from below by the future light cone $x=\pm\varepsilon\mp t$ of the event $(t,x)=(\varepsilon, 0)$ and from above by a hyperbola-like curve $x^{(1)}(t)$. Intersection of the line $x=-\varepsilon+t$ with the curve $x^{(1)}(t)$ determine an instant of time
 \be
 t_c=\frac{a}{\varepsilon}\Big(2-\sqrt{1-v}\Big)\label{tc}
 \ee
at which both supports of the solution $\phi_{2C}(t,x)$ shrink to zero.  It means that $\phi_{2C}(t,x)$ is restricted to a compact region on the Minkowski diagram.  This region is delimited by endpoints $\pm x_c$ where $x_c:=-\varepsilon +t_c$.  The events $(t_c,\pm x_c)$ are marked in Fig.\ref{fig:diag}(a).  Note that $t_c\rightarrow\infty$  for $\varepsilon\rightarrow 0$.

 A partial solution that matches  $\phi_{2C}(t,x)$ at $x^{(1)}(t)=\pm\sqrt{b(t)}$ is symmetric in variable $x$ and negative valued. Imposing matching conditions at  $x=\sqrt{b(t)}$ on a central partial solution $\phi_{3C}(t,x)=F(x+t)+G(x-t)+\frac{t^2}{2}$ we get
 \begin{align}
 &F(\sqrt{b(t)}+t)+G(\sqrt{b(t)}-t)+\frac{t^2}{2}=0,\nonumber\\
 &F'(\sqrt{b(t)}+t)+G'(\sqrt{b(t)}-t)=\frac{a\sqrt{b(t)}}{2(1-v)},\nonumber
 \end{align}
 where $F'(s)=\frac{dF(s)}{ds}$ and similarly for $G'(s)$. Solving these equations we get
\begin{align}
\phi_{3C}(t,x)&=\frac{1}{1+v}\left[H(t+x)+H(t-x)-2H(t_0)\right]+\frac{1}{2}(t^2-t_0^2)\nonumber\\&+\alpha^{(-)}(t^2+x^2-t_0^2)-2\varepsilon\alpha^{(+)}(t-t_0)\label{phi3Creg}
\end{align}
where coefficients $\alpha^{(\pm)}$ have the form
$
\alpha^{(\pm)}=\frac{1}{2}\left(\frac{1}{1-v}\pm\frac{2}{1+v}\right).
$
The function $H(z)$ stands for the integral $H(z):=\int dz \sqrt{c(z)}$, where $c(z)=-vz^2+2\varepsilon z-\varepsilon^2$, and it reads
\begin{align}
H(z)=&\frac{1}{2}\left(z-\frac{\varepsilon}{v}\right)\sqrt{c(z)}-\frac{1}{2}\left(\varepsilon^2\frac{1-v}{(-v)^{\frac{3}{2}}}\ln\left[2a\left(\frac{\varepsilon}{\sqrt{-v}}+z\sqrt{-v}+\sqrt{c(z)}\right)\right]\right).\nonumber
\end{align}
The integration constant is fixed by condition $\phi_{3C}(t_0,0)=0$.
Partial solution \eqref{phi3Creg} is negative valued on the segment $-\sqrt{b(t)}<x<\sqrt{b(t)}$ in the interval of time $t_0<t<t_1$. At $t=t_1$ another zero arises, namely $\phi_{3C}(t_1,0)=0$. Due to complexity of the expression \eqref{phi3Creg} we cannot give an exact solution of this equation. Solving numerically this equation for $a=0.05$ and $\varepsilon=0.05013$ we get $t_1\approx 0.554$.

A validity of our numerical solution is determined by instant of time $t_c$ given by \eqref{tc} and in its central region by a hyperbola-like curve $x^{(2)}(t)$ which describe second zero of $\phi_{3C}(t,x)$ i.e. $\phi_{3C}(t,x^{(2)}(t))=0$. In our example $t_c\approx 0.602$. In order  to get solution valid for $t>t_c$ one has to construct a solution inside the future light cones of of events $(t_c,\pm x_c)$ where $x_c=-\varepsilon +t_c$. Such a solution should match $\phi_{3C}(t,x)$ at $x=\pm [x_c-(t-t_c)]$ and $\phi_{3R/L}(t,x)$ at $x=\pm [x_c+(t-t_c)]$. Since $\phi_{3C}<0$ in vicinity of matching point and $\phi_{3R/L}>0$ then the partial solution inside the future light cone of the event $(t_c,\pm x_c)$ should consist on at least two partial solutions with opposite signs.
Unfortunately, a complexity of expression \eqref{phi3Creg} make impossible obtaining an exact expressions for such solutions. For the same reason we cannot give an exact expression for another partial solution in the central region above hyperbola $x^{(2)}(t)$.

\begin{figure}[h!]
\centering
{\includegraphics[width=0.6\textwidth,height=0.35\textwidth, angle =0]{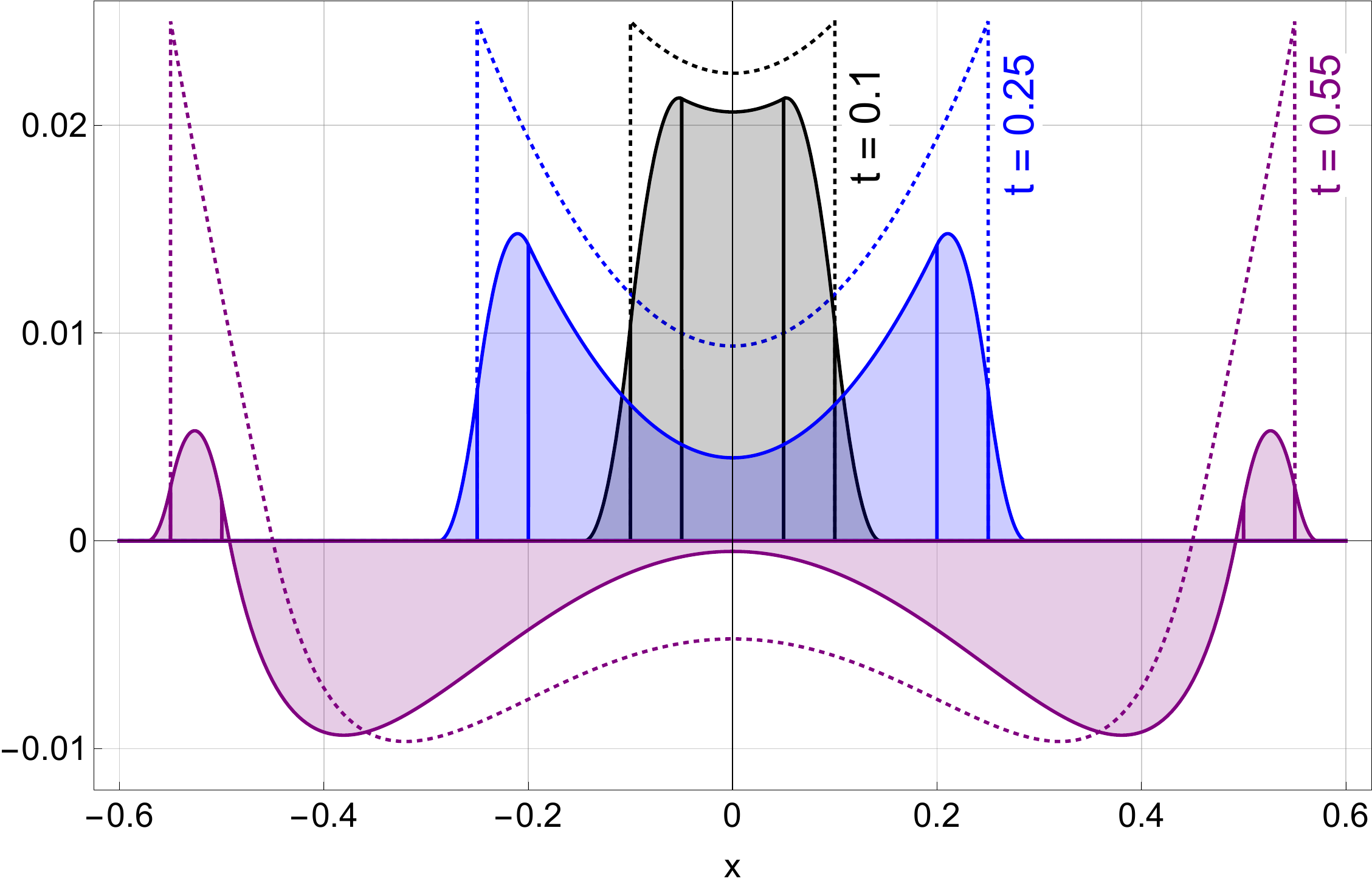}}
\caption{Evolution of a shock-like wave characterized by $a=0.05$ and $\epsilon=2\times 10^{-4}$ and evolution if the exact shock wave for $t=0.1$, $t=0.25$ and $t=0.55$.}
\label{fig:evolution}
\end{figure}

In Fig.\ref{fig:evolution} we show the exact shock-like waves obtained for $a=0.05$ and $\epsilon=2\cdot 10^{-4}$ in three instants of time $t=0.1$, $t=0.25$ and $t=0.55$. The dotted line represent the signum-Gordon field corresponding with an exact shock wave configuration with $a_0=\frac{1}{2}a$. Two external bumps localized in the vicinity of the future light cone of the event $(t,x)=(0,0)$ decrease with time.
\begin{figure}[h!]
\centering
\subfigure[]{\includegraphics[width=0.3\textwidth,height=0.22\textwidth, angle =0]{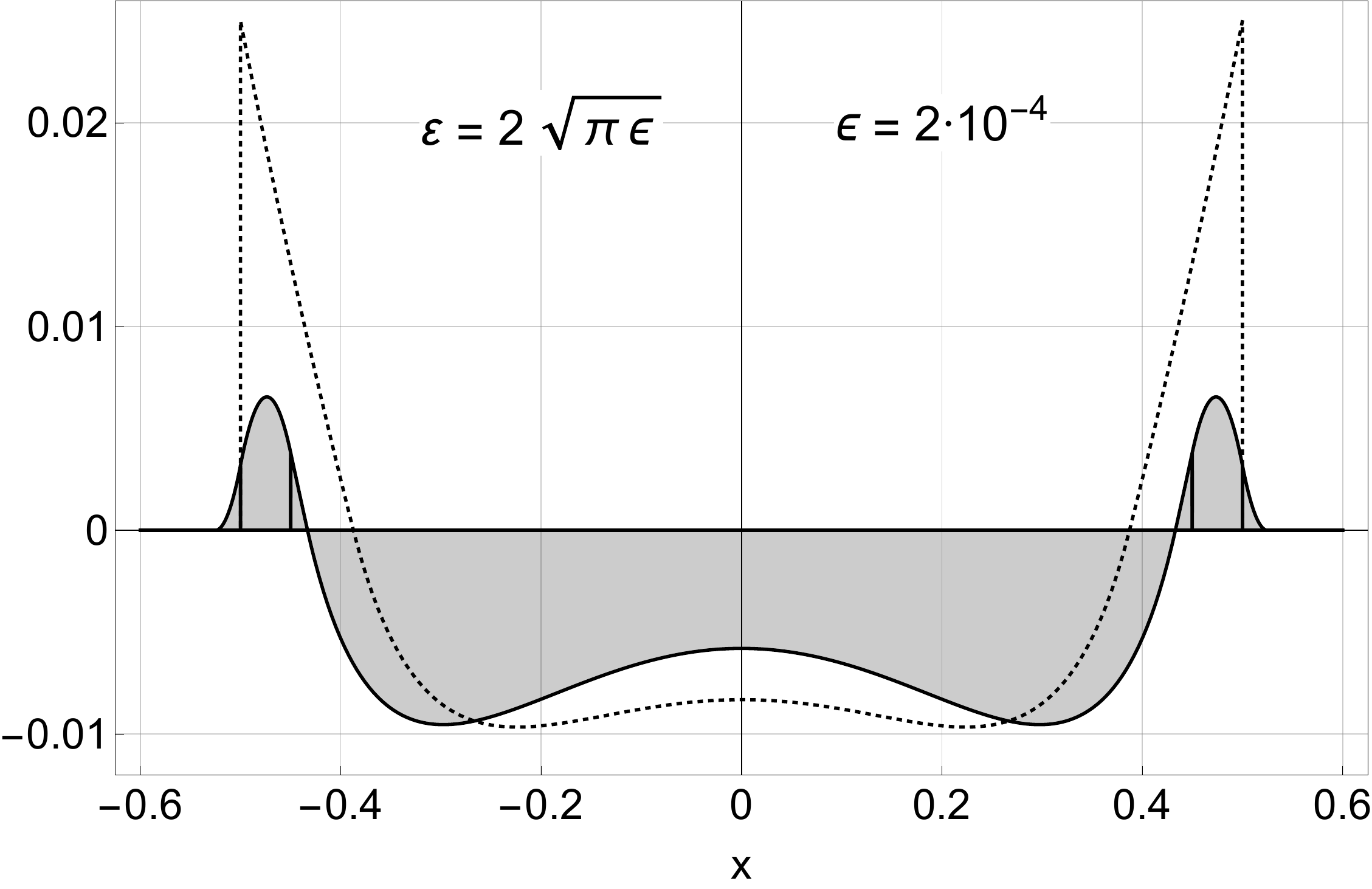}}
\hskip0.2cm
\subfigure[]{\includegraphics[width=0.3\textwidth,height=0.22\textwidth, angle =0]{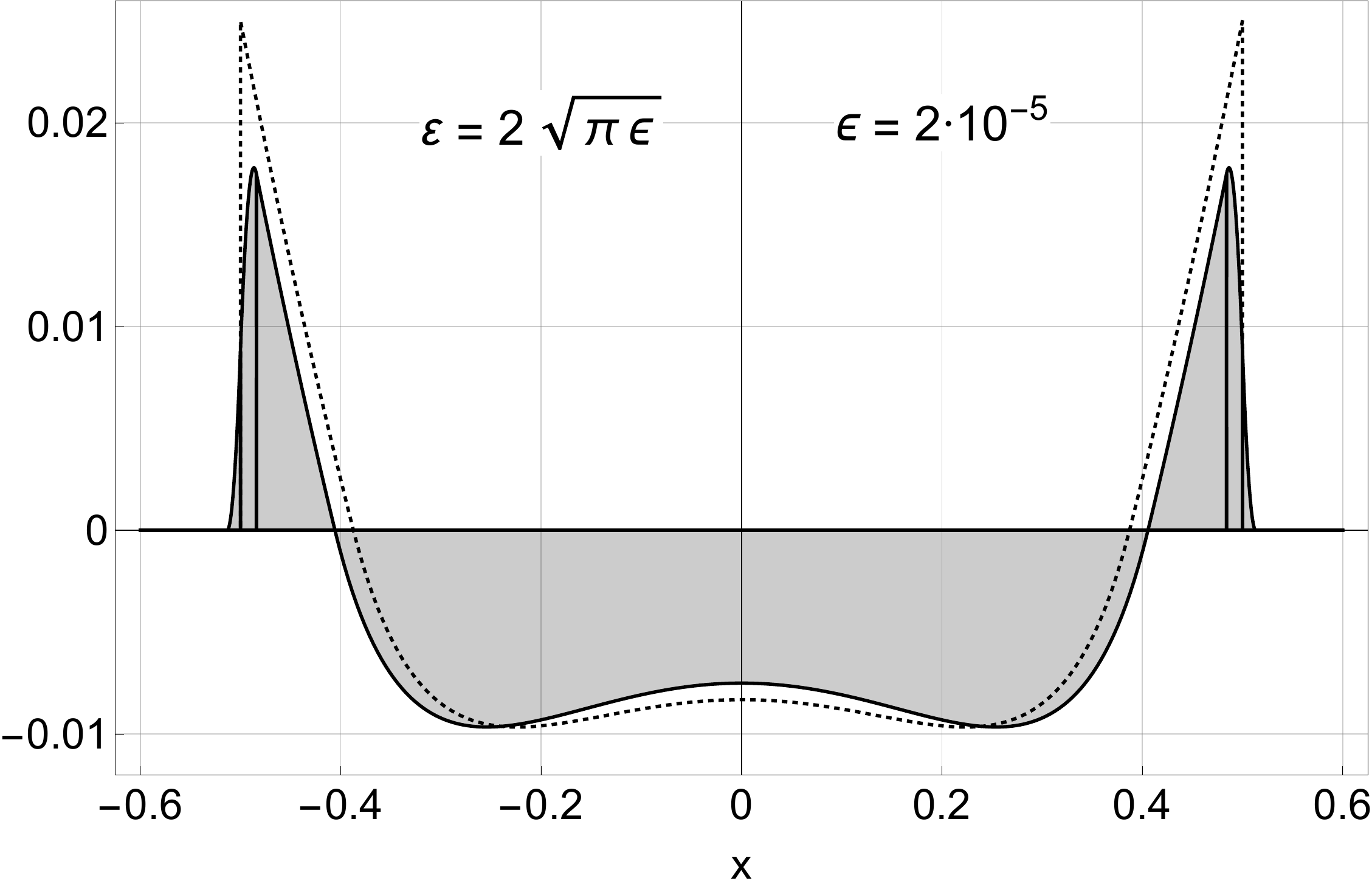}}
\hskip0.2cm
\subfigure[]{\includegraphics[width=0.3\textwidth,height=0.22\textwidth, angle =0]{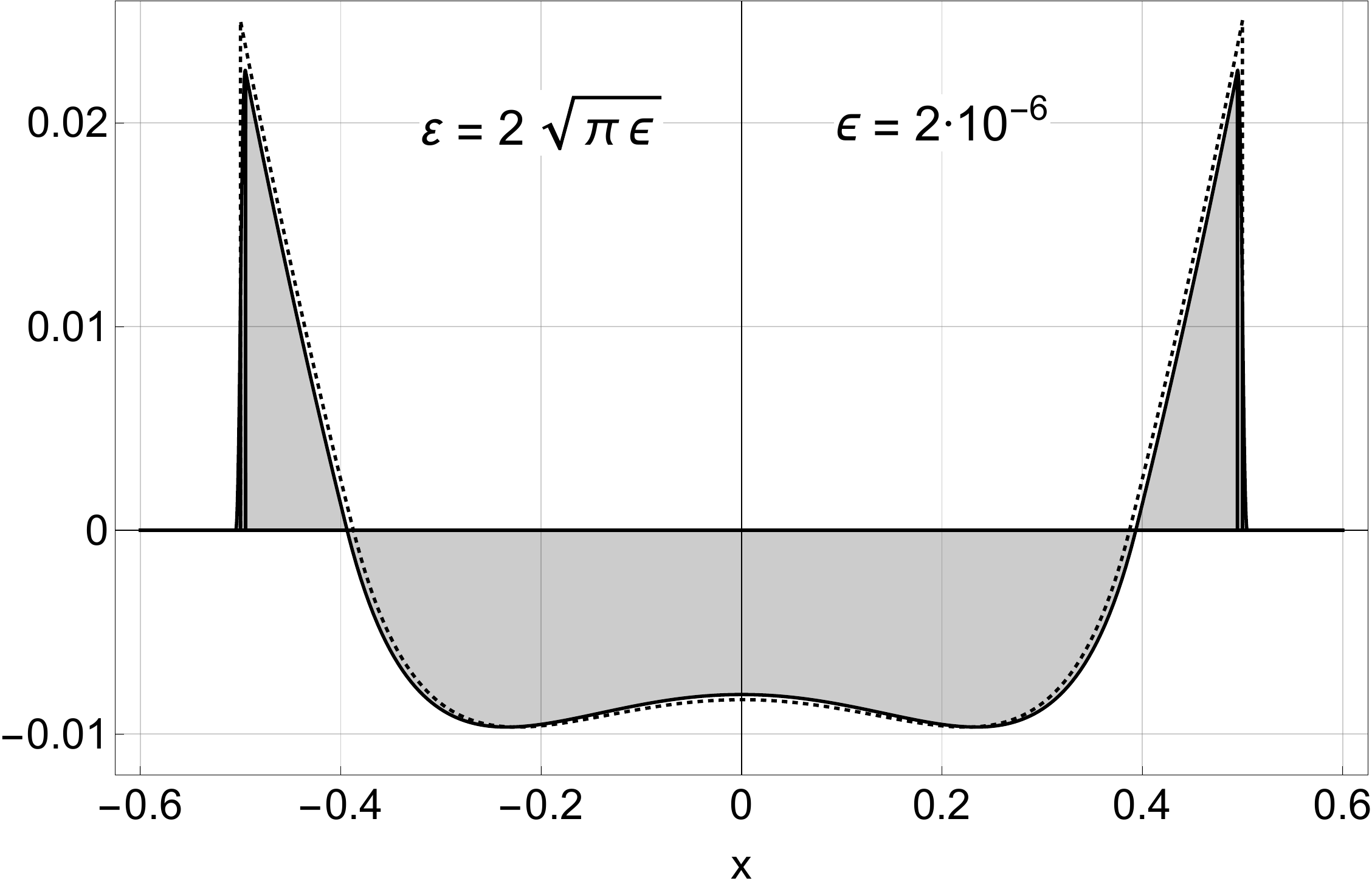}}
\caption{The shock-like wave at $t=0.5$ for $a=0.05$ and (a) $\epsilon=2\cdot 10^{-4}$, (b) $\epsilon=2\cdot 10^{-5}$ and (c) $\epsilon=2\cdot 10^{-6}$. For $\epsilon\rightarrow 0$ the shock wave-like solution (solid curve) tends to exact shock wave solution (dotted curve).}
\label{fig:compare}
\end{figure}

Fig.\ref{fig:compare} shows the exact shock wave  and shock-like wave solutions at $t=0.5$. The regular (without discontinuities) solutions are taken for three different values of the parameter  $\varepsilon=2\sqrt{\pi\epsilon}$ given by $\epsilon=2\cdot 10^{-4}$,  $\epsilon=2\cdot 10^{-5}$ and $\epsilon=2\cdot 10^{-6}$. The shock-like wave solution tends to exact shock wave in the limit $\epsilon\rightarrow 0$.

\begin{figure}[h!]
	\centering
	\subfigure[$ \quad \epsilon = 2 \cdot 10^{-4} $]{\includegraphics[width=0.45\textwidth, height=0.2\textwidth]{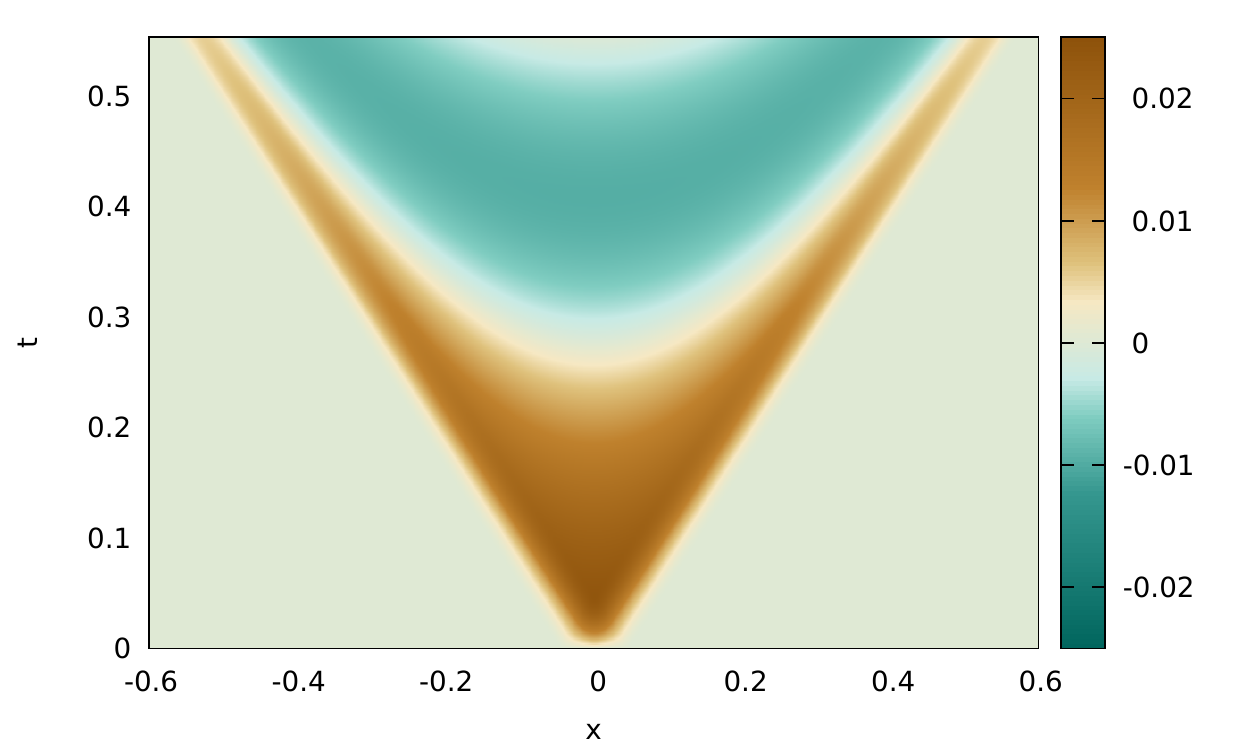}}
	\subfigure[$ \quad \epsilon = 2 \cdot 10^{-4} $]{\includegraphics[width=0.45\textwidth, height=0.2\textwidth]{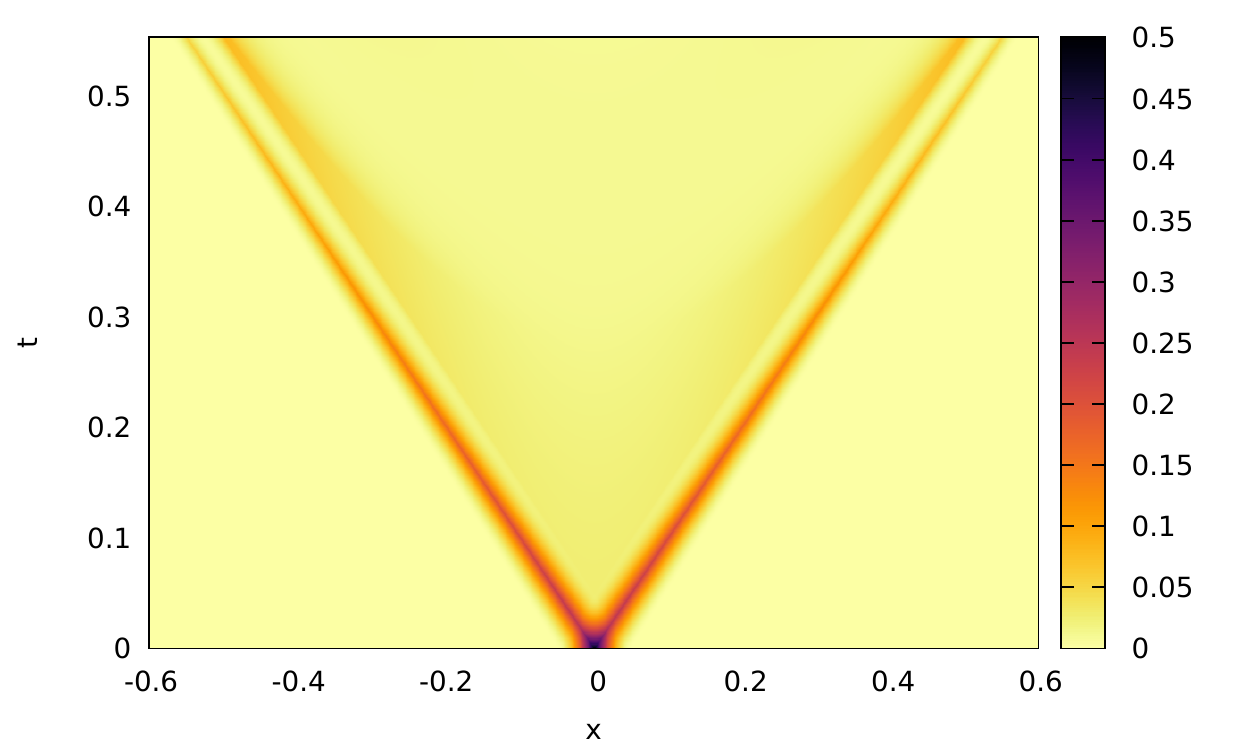}}
	\subfigure[$ \quad \epsilon = 2 \cdot 10^{-5} $]{\includegraphics[width=0.45\textwidth, height=0.2\textwidth]{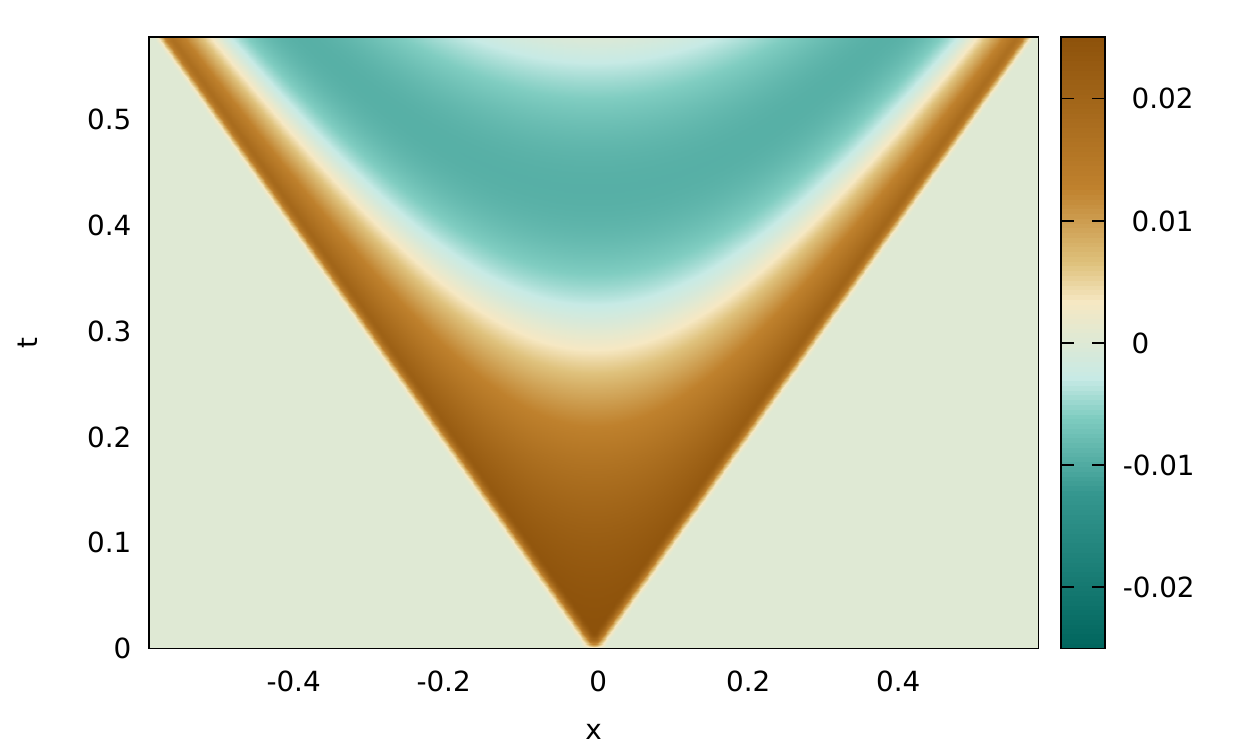}}
	\subfigure[$ \quad \epsilon = 2 \cdot 10^{-5} $]{\includegraphics[width=0.45\textwidth, height=0.2\textwidth]{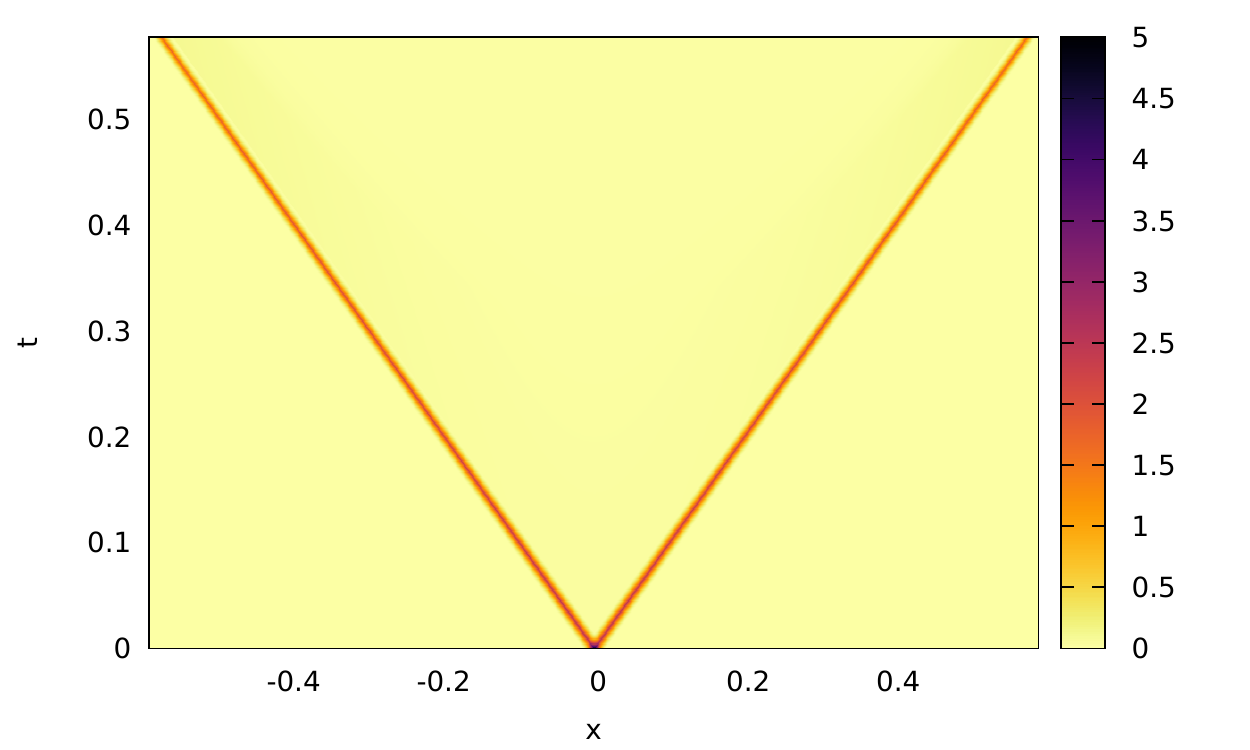}}
	\subfigure[$ \quad \epsilon = 2 \cdot 10^{-6} $]{\includegraphics[width=0.45\textwidth, height=0.2\textwidth]{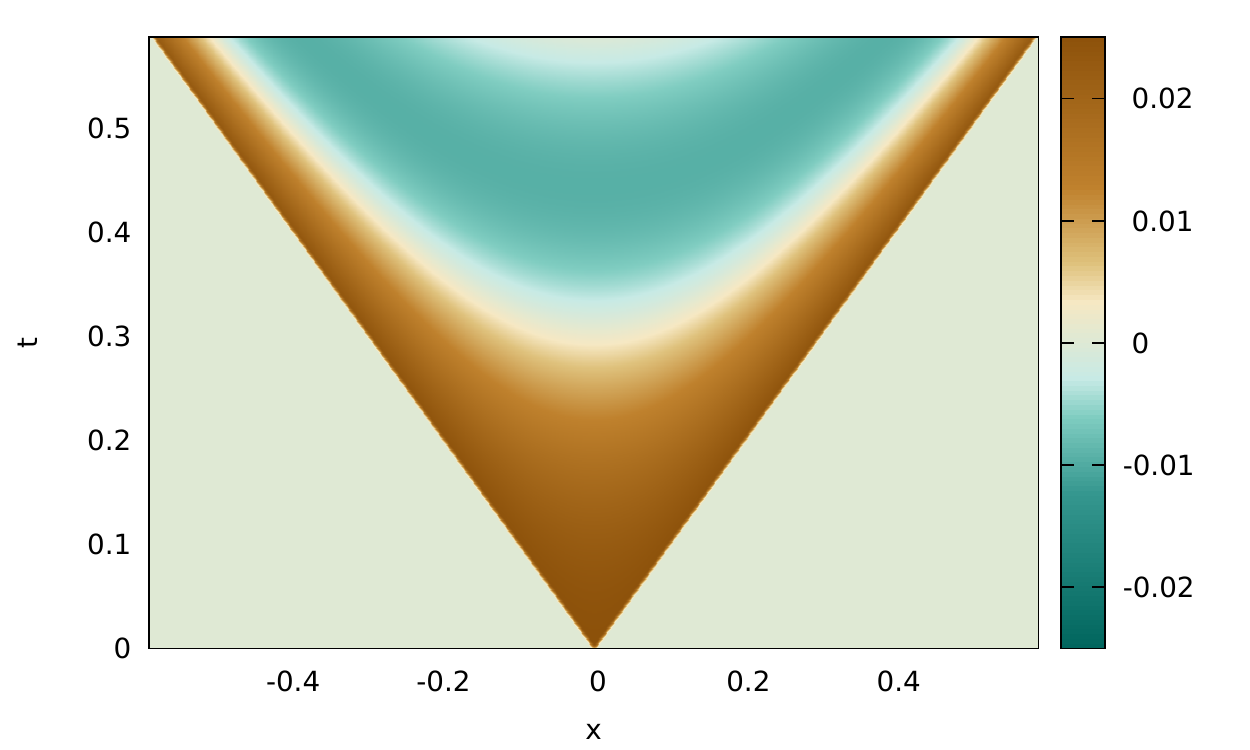}}
	\subfigure[$ \quad \epsilon = 2 \cdot 10^{-6} $]{\includegraphics[width=0.45\textwidth, height=0.2\textwidth]{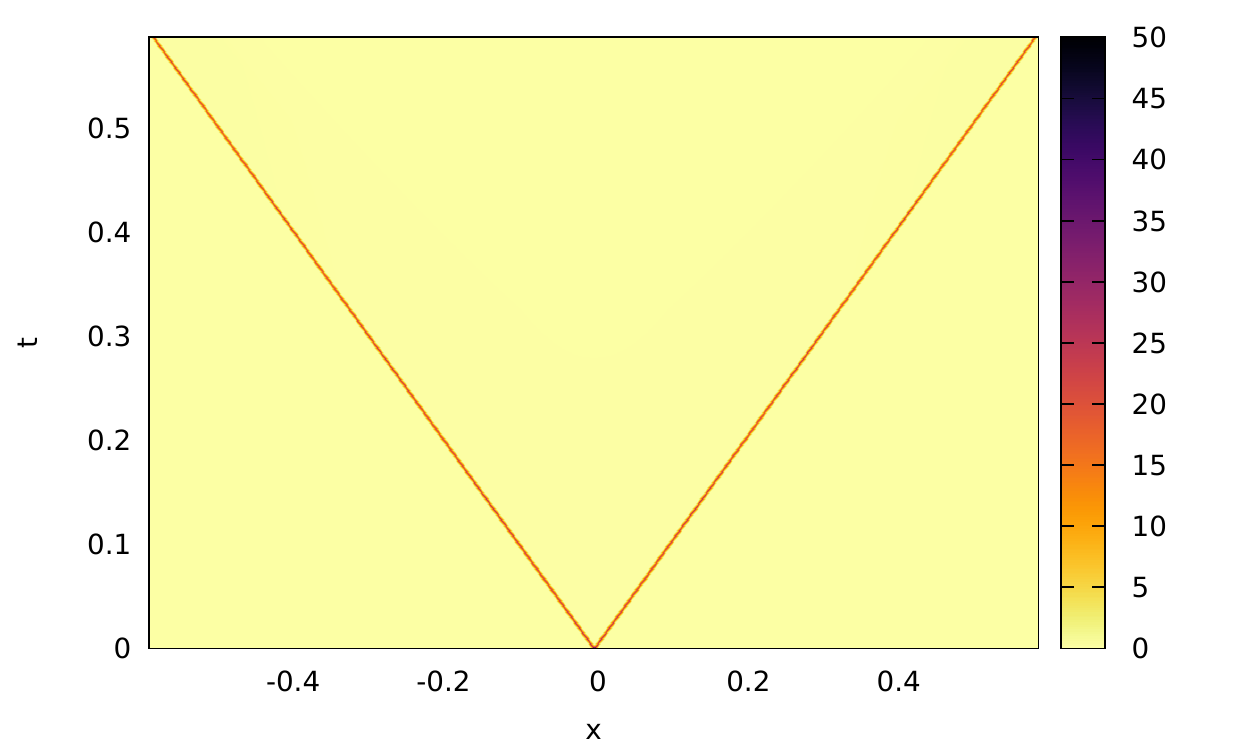}}
	\caption{Field $\phi(t,x)$ (left) and its energy density (right) for $ a = 0.05 $.}
	\label{fig:exact}
\end{figure}

Knowing the exact partial solutions, we can write explicitly expressions for the energy density carried by each piece of the solution:
\begin{align}
	u_{1C}(t,x) =& \frac{(2+v) t^2 - 2 \varepsilon t - v x^2 + \varepsilon ^2}{2 (1+v)^2} \label{eq:exact-u1C} \\
	u_{2C}(t,x) =& \frac{v t^2 - 2 t \varepsilon +(2-v) x^2}{2 (1-v)^2}+\frac{(3-v) \varepsilon ^2}{2 (1-v)^2 (v+1)}\\
	u_{1L}(t,x) =& \frac{t^2+(x+\varepsilon )^2}{2 (1+v)^2}\\
	u_{2L}(t,x) =& \frac{(x + \varepsilon -v t)^2}{(1-v^2)^2}\\
	\begin{split}
		u_{3L}(t,x) =& \frac{1}{2 (1-v^2)^2}\Big[-v\qty(1-v^2)\qty(t^2 - x^2) + \qty(1+v^2)\qty(t^2  + x^2) \\& -2 t ((1 + v^2) \varepsilon +2 v x ) +2 \qty(1+(2-v)v) x \varepsilon +2 \varepsilon ^2\Big]
	\end{split}.
	\label{eq:exact-u3L}
\end{align}
We do not write $ u_{3C}(t,x) $ because its expression is too complicated to be analytically obtained. However, we can use the finite differences method to take the derivatives of $ \phi_{3C}(t,x) $ necessary to calculate
its energy density.

The expressions \ref{eq:exact-u1C}--\ref{eq:exact-u3L} and the numerical data for $ u_{3C}(t,x) $ were used to plot the energy density as a color gradient for $ 0 \leq t \leq t_1 $ (Fig. \ref{fig:exact}, right-hand side).

Looking closely to Fig.\ref{fig:exact} we can see the energy density getting dimmer as the solution evolves in time. Such effect is particularly visible for larger values of $ \epsilon $ and gets less noticeably as $ \epsilon $ decreases. This behavior suggests that the outer regions of the solution be as a reservoir of energy, feeding the expansion of the inner structures (shock wave like).

The partial solutions $ \phi_{2C} $ and $ \phi_{3C} $ corresponds to partial solutions of the exact shockwave. Therefore, $ \phi_{C}, \phi_{1L}, \phi_{2L}, \phi_{3L} $ (and the corresponding right-side solutions) are related to our approximation of the Dirac delta by a finite function. When $ \epsilon \rightarrow 0 $, these solutions reduce to a discontinuity in the field. The color gradient plots suggests that this partial solutions loose energy over time. We can examine this claim closer by integrating the energy densities and obtain the total energy as a function of time:

\begin{align}
	E_{1C}(t) &=
	\begin{cases}
		\frac{t \left(2 t^2 (v+3)-6 t \varepsilon +3 \varepsilon ^2\right)}{3 (v+1)^2} & \quad \text{if }\quad 0 \leq t \leq \frac{\varepsilon}{2}\\
		\frac{(\varepsilon -t) \left(2 t^2 (v+3)+2 t (v-3) \varepsilon -(v-3) \varepsilon ^2\right)}{3 (v+1)^2} & \quad \text{if }\quad \frac{\varepsilon}{2} < t \leq \varepsilon
	\end{cases}\\
	E_{1L}(t) = E_{1R}(t) &= \frac{-8 t^3+6 t^2 \varepsilon -3 t \varepsilon ^2+\varepsilon ^3}{6 (v+1)^2}\\
	E_{2L}(t) = E_{2R}(t) &=
	\begin{cases}
		-\frac{t^3 (v-1)}{3 (v+1)^2} & \quad \text{if }\quad 0 \leq t \leq \frac{\varepsilon}{2}\\
		-\frac{((1+v)t-\varepsilon )^3}{3 \left(v^2-1\right)^2} & \quad \text{if }\quad t > \frac{\varepsilon}{2}
	\end{cases}\\
	E_{3L}(t) = E_{3R}(t) &=
	\begin{cases}
		\frac{(2 t-\varepsilon ) \left(2 t^2 [v (v (v+2)-1)+2]+t [v (v (v-7) +5)-7] \varepsilon -[v (v(v-4) +5)-4] \varepsilon ^2\right)}{6 \left(v^2-1\right)^2} & \quad \text{if } \quad\frac{\varepsilon}{2} \leq t \leq \varepsilon \\
		\frac{\varepsilon  \left(6 t^2 (v+1)^2+3 t [v(v-2) -5] (v+1) \varepsilon -[v (v (v+2)-7)-10] \varepsilon ^2\right)}{6 \left(v^2-1\right)^2} & \quad \text{if }\quad t > \varepsilon.
	\end{cases}
\end{align}

Considering the intervals of time where each solution is valid, we can sum up the energy contained in the outer regions of the field. That is, the energy contained in the regions that are reduced to discontinuities when $ \epsilon $ tends to zero.

\begin{figure}[h!]
	\includegraphics[width=0.5\textwidth]{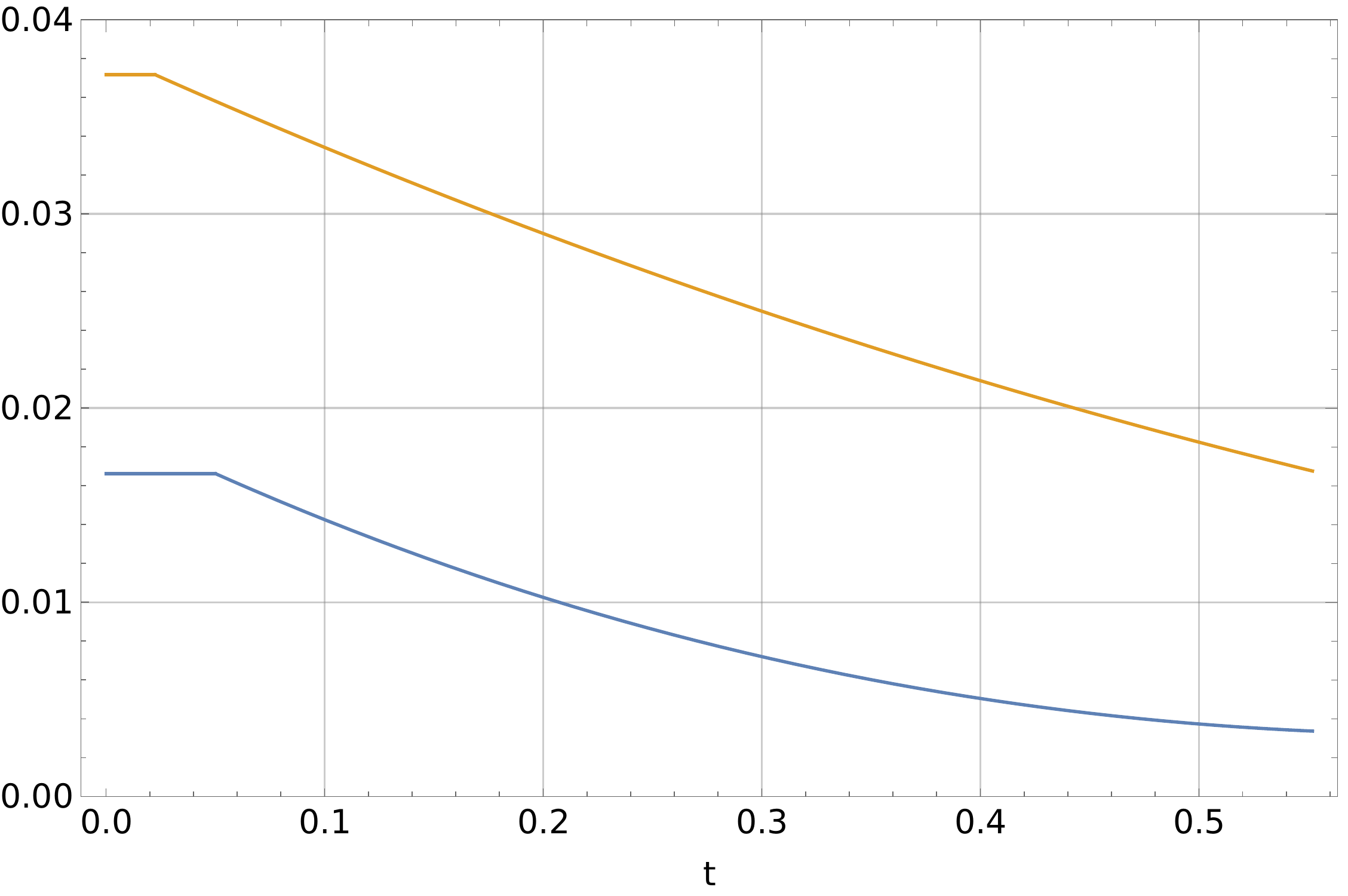}
	\caption{Total energy carried by the partial solutions $\phi_{1C}$, $\phi_{1L}$, $\phi_{1R}$, $\phi_{2L}$, $\phi_{2R}$, $\phi_{3L}$ and $\phi_{3R}$ for $ \epsilon = 2 \cdot 10^{-4} $ (bottom) and $ \epsilon = 4 \cdot 10^{-5} $ (top).}
	\label{fig:energy-border}
\end{figure}
Looking at Fig.~\ref{fig:energy-border} we see that the energy in the outer regions is initially constant
\begin{align}
E_{1C}(t)+E_{1L}(t)+E_{1R}(t)+E_{2L}(t)+E_{2R}(t)=\frac{a^2}{3\varepsilon}&=\frac{a^2}{6\sqrt{\pi\epsilon}},\quad \text{if} \quad 0 \leq t \leq \frac{\varepsilon}{2},\nonumber\\
E_{1C}(t)+E_{3L}(t)+E_{3R}(t)+E_{2L}(t)+E_{2R}(t)=\frac{a^2}{3\varepsilon}&=\frac{a^2}{6\sqrt{\pi\epsilon}},\quad \text{if} \quad \frac{\varepsilon}{2} < t \leq \varepsilon,\nonumber
\end{align}
and later it decreases approximately linearly for small $\varepsilon$. This decreasing of energy originates in appearance of inner solution $\phi_{2C}$ and then $\phi_{3C}$. It can be expected that when the energy gets sufficiently small it can no longer feed the expansion of the shockwave. Therefore the shockwave breaks down and radiates.


\subsection{Shock waves and scattering of oscillons}
One of the most characteristic properties of a radiation generated from initial data \eqref{inireg} is an amazing similarity between their patterns and the patterns formed by radiation released during the process of scattering of oscillons.
\begin{figure}[h!]
\centering
\subfigure[$\quad V=0.98,\quad \alpha=0.648$]{\includegraphics[width=0.44\textwidth,height=0.25\textwidth, angle =0
								]{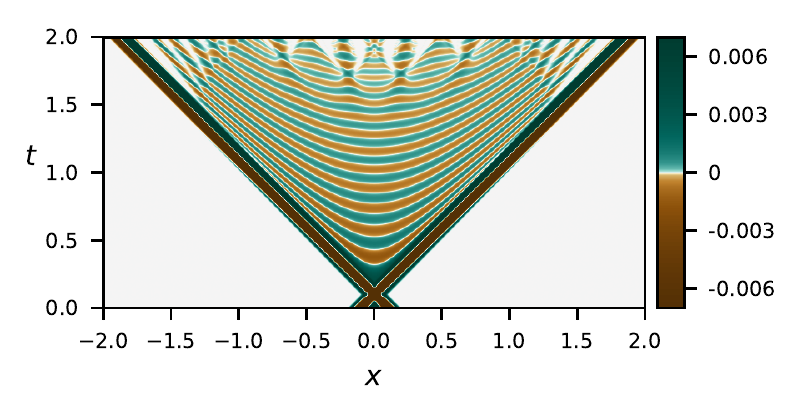}}
\subfigure[$\quad a=0.005,\quad \epsilon=1.747 \cdot10^{-5}$]{\includegraphics[width=0.44\textwidth,height=0.25\textwidth, angle =0
								]{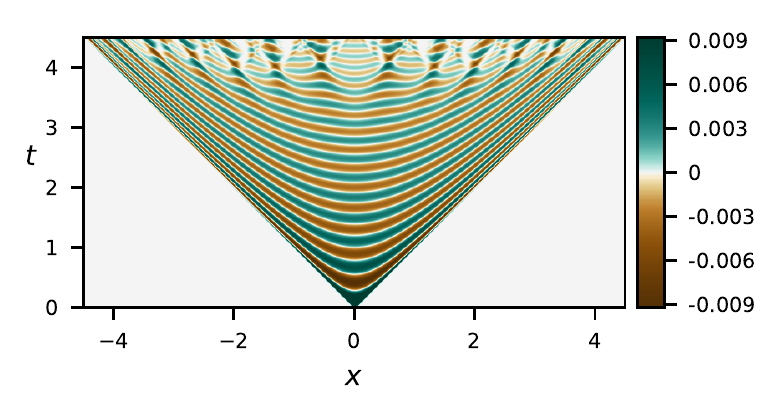}}
\subfigure[$\quad V=0.93,\quad \alpha=0.680$]{\includegraphics[width=0.44\textwidth,height=0.25\textwidth, angle =0
								]{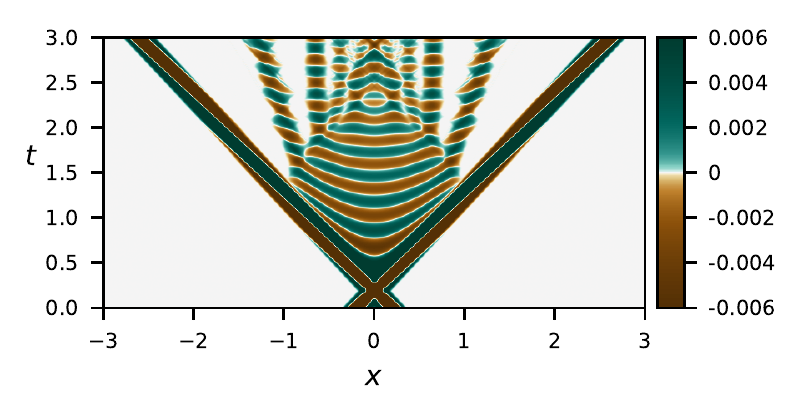}}
\subfigure[$\quad a=0.005,\quad\epsilon=2.288\cdot 10^{-4}$]{\includegraphics[width=0.44\textwidth,height=0.25\textwidth, angle =0
								]{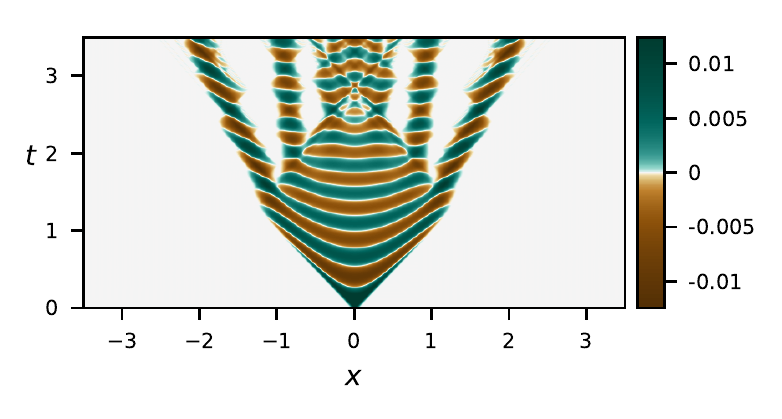}}
\subfigure[$\quad V=0.74,\quad \alpha=0.089$]{\includegraphics[width=0.44\textwidth,height=0.25\textwidth, angle =0
								]{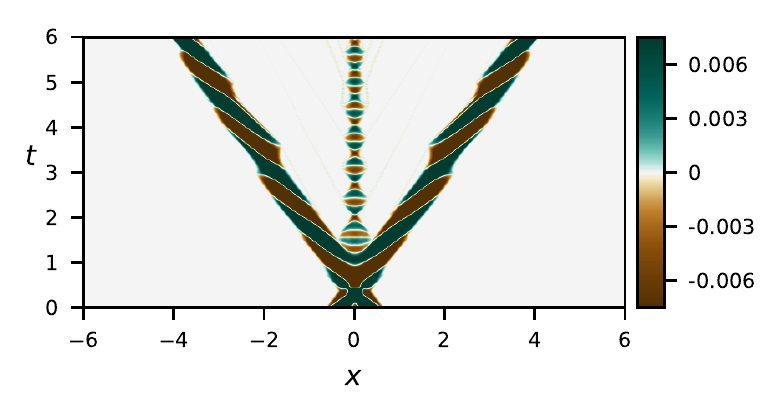}}
\subfigure[$\quad a=0.005,\quad \epsilon=3.0959\cdot 10^{-2}$]{\includegraphics[width=0.44\textwidth,height=0.25\textwidth, angle =0
								]{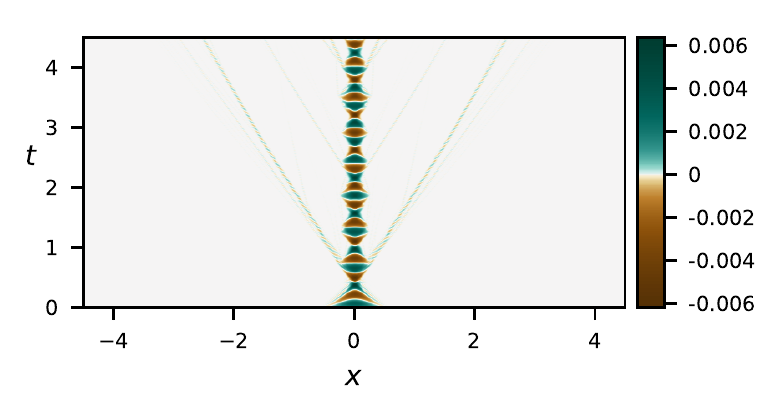}}
\caption{Scattering od symmetric oscillons (left) and evolution of initial data \eqref{inireg} (right).}
\label{fig:compare2}
\end{figure}
The problem of scattering of oscillons has been reported in Ref. \cite{scattering}. The initial configuration contains two exact oscillons with compact supports that  touch each other at $t=0$. Such oscillons move in front of each other with equal speeds in the laboratory reference frame. In order to simplify considerations we take only the initial configurations which are symmetric under spatial reflections $x\rightarrow -x$. The scattering process leads to the emergence of two main oscillons and production of the radiation (smaller oscillons waves, etc). This radiation is mainly concentrated  in the central region of the spacetime diagram. There are two parameters which we use to get different initial configurations: speeds of initial oscillons $V$ and their phase $\alpha$. Dependently on the value of these parameters (on the form of the initial oscillons) we get a variety of different patterns of the radiation.

In Fig.\ref{fig:compare2} we show three examples of the evolution of the signum-Gordon field. The magnitude of the field is represented by a gradient color. Subfigures (a), (c), (e) show scattering of two oscillons in dependence on their initial speed $V$ and the phase $\alpha$ whereas subfigures (b), (d), (f) show evolution of some initial field configurations given by \eqref{inireg} (Gaussian case) with different values of $\epsilon$. The left and right figures look amazingly similar taking into account that there is no direct relation between their initial conditions. The presence of a radiation in the scattering process means that two emerging (leading) oscillons has less energy than the incoming ones. The difference of the energy is carried by the radiation. Having in mind that the appearance a shock wave solution requires a delta-like initial field configuration we can speculate that such a configuration of the field could be produced shortly after collision (at the moment when two main outgoing oscillons arise). This field configuration would give rise to the shock wave solution in the similar way as the initial data \eqref{inireg} does. In this scenario
the role of a scattering process is limited to generation of a field configuration which furthermore develops a shock wave-like form. It explains to some extent the universal character of the patterns formed by radiation of the signum-Gordon field.

\section{Conclusions}
We have presented some considerations concerning the problem of a collapse of shock-like  wave solutions in the signum-Gordon model. This phenomenon  was observed previously in collision between two oscillons. Since the oscillons dominate the radiation spectrum of the model and they collide frequently the decay of shock waves  is an efficient mechanism of production of small size oscillons.

In the first part of the  paper we looked in more detail at the exact shock wave solution.  We managed to establish initial condition for  exact shock waves. Namely, a suitable  initial field configuration contains the field which vanishes everywhere and its time derivative is proportional to the Dirac delta. The support of the wave is localized inside the light cone including the cone itself. We have shown that the energy of the solution inside the light cone (excluding the cone)  increases linearly with time. This result may look strange when confronted with the fact that the signum-Gordon model conserves the energy. In fact there is no inconsistency between these two facts. The total energy of the shock waves includes also the gradient energy associated with discontinuities of the field at the light cone. This energy is clearly infinite. Thus the existence of the exact shock waves requires a continuous transfer of the energy from discontinuities to the region inside the light cone.

Next we have looked at the systems with finite total energy. Taking regular $\delta$-like initial profile of $\partial_t\phi$ we got a numerical solution which initially looks very similar to exact shock wave solution. The finiteness of the gradient energy of the field close to the light cone did not allow for existence of the wave for arbitrary long times. We found that the wave begin to collapse (decay) into oscillons. The decay starts earlier for configurations with lower the gradient energy (bigger $\epsilon$).

The numerical solutions obtained for initial configurations that differ by value of $\epsilon$ are very similar to solutions obtained in the process of scattering of exact oscillons.  Our analysis of shock waves allows for better understanding how oscillons are produced in collision of two incoming oscillons: the outgoing oscillons have less energy than the incoming ones what leads to production of waves that eventually decay into oscillons.

The observed way of production of oscillons in a decay of shock-like wave solution is also expected in other models with V-shaped potentials in the limit of small amplitudes of the field.  Our preliminary results from scattering of compact kinks in double well potential confirm this steatment.

\section*{Acknowledgements}
The authors would like to thank H. Arod\'z, A. Wereszczy\'nski and W. J. Zakrzewski for  discussions and comments. FMH is supported by  CNPq Scholarship and JSS by CAPES Scholarship. This study was financed in part by the Coordena\c c\~ao de Aperfei\c coamento de Pessoal de N\'ivel Superior a Brasil (CAPES) a Finance Code 001.

\end{document}